\documentclass[journal]{IEEEtran}

\usepackage{epsf}
\usepackage{float}
\usepackage{graphicx}
\DeclareGraphicsExtensions{.png}
\usepackage{caption}
\usepackage{subcaption}
\usepackage{xcolor}
\usepackage{amsmath} \interdisplaylinepenalty=2500
\usepackage{amsfonts}
\usepackage{amssymb}
\usepackage{array}
\usepackage{cuted}

\newcommand{\dul}[1]{\underline{\underline{#1}}}

\newcommand{\bea}{\begin{eqnarray}}

\newcommand{\eea}{\end{eqnarray}}

\newcommand{\rmc}{{\rm c}}

\newcommand{\rmd}{{\rm d}}

\newcommand{\rmEMI}{{\rm EMI}}

\newcommand{\rmj}{{\rm j}}

\newcommand{\cE}{{\cal E}}

\newcommand{\cF}{{\cal F}}

\begin{document}



\title{Correlation and Spectral Density Functions in Mode-Stirred Reverberation -- III. Measurements
}
\author{
{Luk R. Arnaut 
and John M. Ladbury
}
}


\maketitle



\begin{abstract}
Experimental auto- and cross-correlation functions and their corresponding spectral density functions are extracted from measured sweep data of mode-stirred fields. These are compared with theoretical models derived in part I, using estimated spectral moments from part II. The second-order Pad\'{e} approximant based model accounts for the main features of the spectral density function, including its slope near stir DC, corner frequency, stir DC-to-Nyquist level drop, and asymptotic spectral density. Ensemble averaging across secondary tune states offers a reduction of spectral bias and RMS spectral fluctuation, compared to spectral densities for individual stir sweeps or their concatenation. Periodogram- and correlation-based methods produce near-identical results. Distinctive theoretical features between power-based vs. field-based spectral densities are experimentally verified. Interchanging the roles of stirrer and tuner demonstrates the effect of stir efficiency on correlation and spectral density. The spectral characterization allows for stirrer diagnostics, which is demonstrated through detection and identification of EMI caused by mains power harmonics in the measured stir spectrum at low frequencies.
\end{abstract}


\section{Introduction\label{sec:intro}}
In this part III article, measurement data obtained in a dual-paddle mechanically mode-stirred reverberation chamber (MSRC) are used to extract empirical correlation functions (CFs) and spectral density functions (SDFs). 
The focus is on overmoded operation (CW operating frequencies $f \geq 2$ GHz), enabling a validation of theoretical results derived in \cite{arnaACFSDF_pt1},  \cite{arnaACFSDF_pt2}.

Previously \cite{arnalocavg}, power autocorrelation functions (ACFs) and SDFs in a mode-tuned chamber were obtained experimentally and modelled empirically, in a simplified way, based on scalar measurements using a power sensor and applying high-order exponential curve fitting.
Nowadays, the availability of high-speed high-capacity vector spectrum or network analyzers (VSAs/VNAs) allows for extending that work significantly. Vector measurements do not require an assumption of idealized complex circularity of the underlying received stirred field $E(\tau)$ in \cite{arnalocavg}, while also enabling the characterization of cross-correlation functions (CCFs) and cross-spectral density functions (CSDFs) for in-phase/quadrature (I/Q) field components.
In addition, the high sampling rate of VSAs/VNAs offers insight into limitations of sampled sweeps for capturing the actual (i.e., time-continuous) physical field evolution, including stir noise data and its correlation properties \cite{arnaACFSDF_pt2}. 

A major motivation for modelling and validating ACFs and SDFs is to build an accurate model of the stir process in support of a theory of mode stirring in continuous time (dynamic EM environment), i.e., beyond discrete mode tuning. Here, it will be demonstrated that mode stirred operation enables certain EMI effects to be detected that would normally not be noticeable in typical mode-tuned operation that exhibits a relatively large dwell time per tune state.

In this third part, the theoretical expressions for the CFs and SDFs for both fields and power as input quantities in \cite{arnaACFSDF_pt1} are compared and validated against experimental data to demonstrate the feasibility of the models and to assess their accuracy. 
The correlation structure is characterized by spectral moments, which are calculable independently \cite{arnaACFSDF_pt2}. 
All temporal and spectral processing is performed in ``raw'' format, i.e., without temporal nor spectral windowing for tapering of finite data sets, except for the truncation to one full stir period \cite[sec. IV.A]{arnaACFSDF_pt2}, in order to avoid biasing the mean and (co)variance. 

The following notational conventions are adopted. $m$ overdots indicate $m$th-order differentiation with respect to the dependent variable, being either the temporal continuous stir state $\tau$ or the stir frequency $\varpi$ with sampled values $\tau_n$ and $\varpi_k$, respectively. Single- and double-primed quantities refer to either the real (in-phase) or imaginary (quadrature) part of the complex electric field $E(\tau) = E^\prime(\tau) + \rmj E^{\prime\prime}(\tau)$ with an assumed $\exp(\rmj \varpi\tau)$ stir dependence, or its complex CFs. 
We refer to auto- and cross-correlations or -spectra as relating to the I- and Q-components of the complex $E$. 
Alternatively, one can obtain the complex ACF of $E$ with the complex conjugate of itself, $E^*$.
For economy of notation, $E^{\prime(\prime)}$ denotes both $E^\prime$ and $E^{\prime\prime}$ as separate cases combined within a single expression, and similarly for $\lambda^{\prime(\prime)}_i$. As in \cite{arnaACFSDF_pt1}, primed spectral moments $\lambda^{\prime(\prime)}_i$ already imply normalization by the field variance, $\lambda_0$.

\section{Experimental Set-Up}
Details of the chamber, measurement configuration and instrumentation were described in \cite[sec. V-A.]{arnaACFSDF_pt2}.
The measurements reported here refer to the MSRC  including an RF amplifier. Except for sec. \ref{sec:interchange}, the `large' short wide paddle is used in stirred mode (continuous rotation) while the `small' tall narrow paddle is operated in tuned mode (stepped rotation). In sec. \ref{sec:interchange}, the roles of both paddles are interchanged.

\section{Correlation: ACF and CCF}
\subsection{ACF}
For a wide-sense stationary (WSS) quasi-circular (QC) field $E(\tau)=E^\prime(\tau) +\rmj E^{\prime\prime}(\tau)$ at time lag $\tau  =t-t_0$, its (real) ACF $\rho^\prime_{E}(\tau)$ and CCF $\rho^{\prime\prime}_{E}(\tau)$ can be combined to the complex ACF
\begin{align}
\rho_{E}(\tau) &\equiv \rho^\prime_E(\tau) + \rmj \rho^{\prime\prime}_E(\tau) = {\langle E(0) E^*(\tau) \rangle} / {\sigma^2_{E}}
\label{eq:def_rho_p}
\end{align}
where $\sigma^2_{E} \simeq \sigma^2_{E^\prime} + \sigma^2_{E^{\prime\prime}} \simeq 2 \sigma^2_{E^{\prime(\prime)}}$ and 
\begin{align}
\rho^\prime_{E}(\tau) &\simeq 
\rho_{E^{\prime(\prime)}}(\tau) 
= {\langle E^{\prime(\prime)}(0) E^{\prime(\prime)}(\tau) \rangle} / {\sigma^2_{E^{\prime(\prime)}}}
\label{eq:rho_p}\\
\rho^{\prime\prime}_{E}(\tau) &\simeq 
-\rho_{E^\prime,E^{\prime\prime}}(0,\tau) 
= - {\langle E^\prime(0) E^{\prime\prime}(\tau) \rangle} / (\sigma_{E^{\prime}} \sigma_{E^{\prime\prime}})
.
\label{eq:rho_pp}
\end{align}
For $N_s$ data samples $\{E_n\}^{N_s-1}_{n=0}$ at $\tau_n = n\Delta \tau =t_n-t_0$, this yields the $N_s\times N_s$ covariance matrix $\dul{R}_E$ of $E$. 

Alternatively, the CFs can also be obtained starting from the SDFs based on the periodogram, using the Fourier transform $\cE(\varpi)$ of the stir data $E(\tau)$, viz., \cite[eq. (32)]{arnaACFSDF_pt1}
\begin{align}
&g_E(\varpi) \equiv g^\prime_E(\varpi) + \rmj g^{\prime\prime}_E(\varpi)
\nonumber\\
&= \left [ \langle \left (\cE^\prime(\varpi) \right )^2 \rangle
+ \langle \left (\cE^{\prime\prime}(\varpi) \right )^2 \rangle 
+ \rmj 2 \langle \cE^\prime(\varpi) \cE^{\prime\prime}(\varpi) \rangle \right ] / \sigma^2_E
.
\label{eq:periodgram_p}
\end{align}
For $N_s$ samples at times $\{\tau_n\}$ and $N_s$ stir frequencies $\{\varpi_k\}$ ($n,k=0,\ldots,N_s-1$), the periodogram can be computed efficiently and more rapidly with the aid of an FFT algorithm than a calculation of $\dul{R}_E$ based on (\ref{eq:def_rho_p}), particularly when $N_s$ is large. 
Thus, following the Einstein-Wiener-Khinchin (EWK) theorem, the CFs can be efficiently obtained as \cite[eq. (33)]{arnaACFSDF_pt1}
\begin{align}
\{ E(\tau_n) \} \stackrel{{\rm FFT}} {\longrightarrow} 
\{ \cE(\varpi_k) \} \stackrel{(\ref{eq:periodgram_p})}{\longrightarrow} \{g^{\prime(\prime)}_{E}(\varpi_k\} )\stackrel{{\rm IFFT}}{\longrightarrow} 
\{\rho^{\prime(\prime)}_E(\tau_n) \}
.
\end{align}

\begin{figure}[htb] 
\begin{center}
\begin{tabular}{c}
\vspace{-0.5cm}\\
\hspace{-0.6cm}
\includegraphics[scale=0.65]{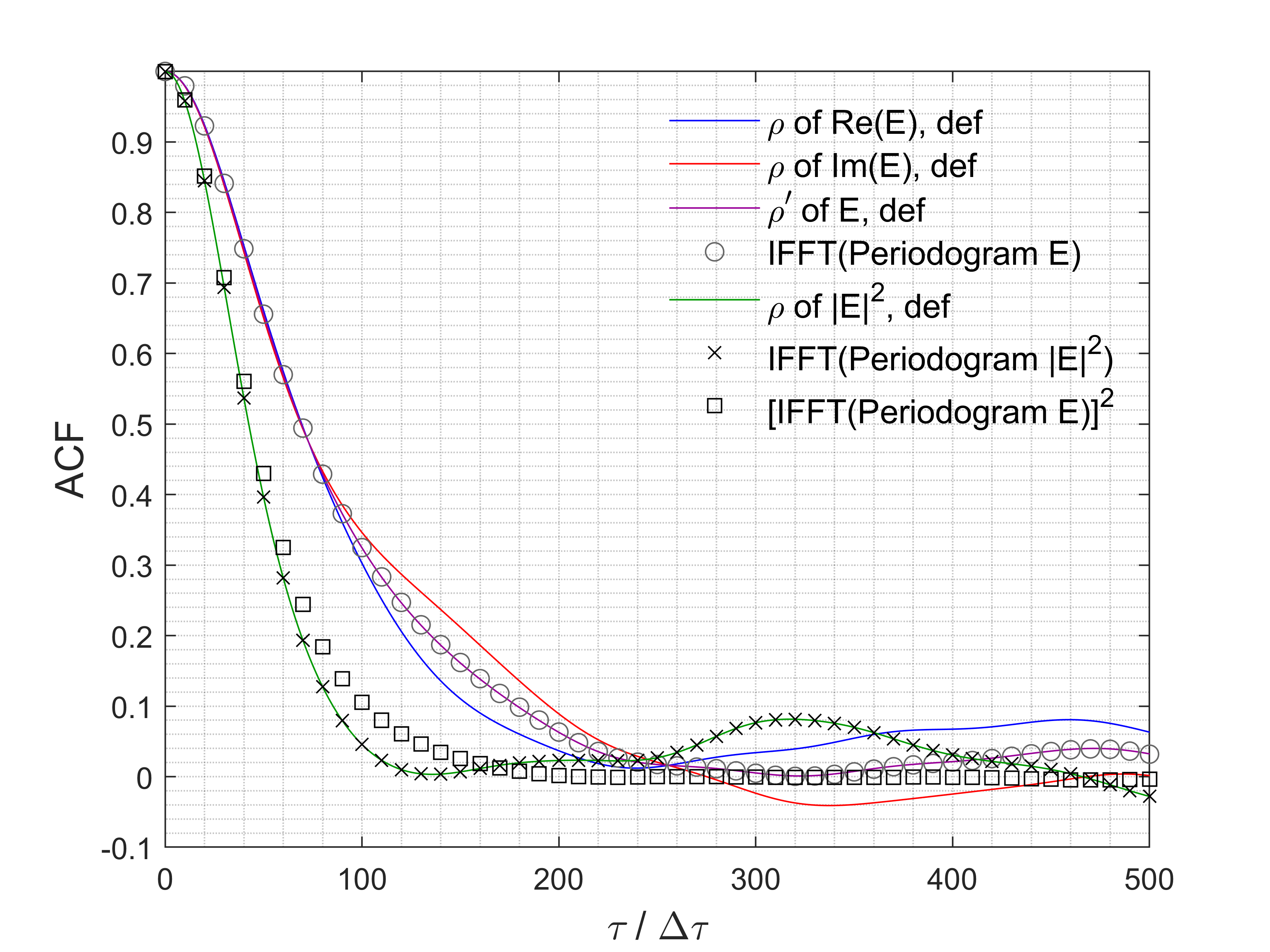}\\
\vspace{-1cm}\\
\\
(a)\\
\hspace{-0.8cm}
\includegraphics[scale=0.65]{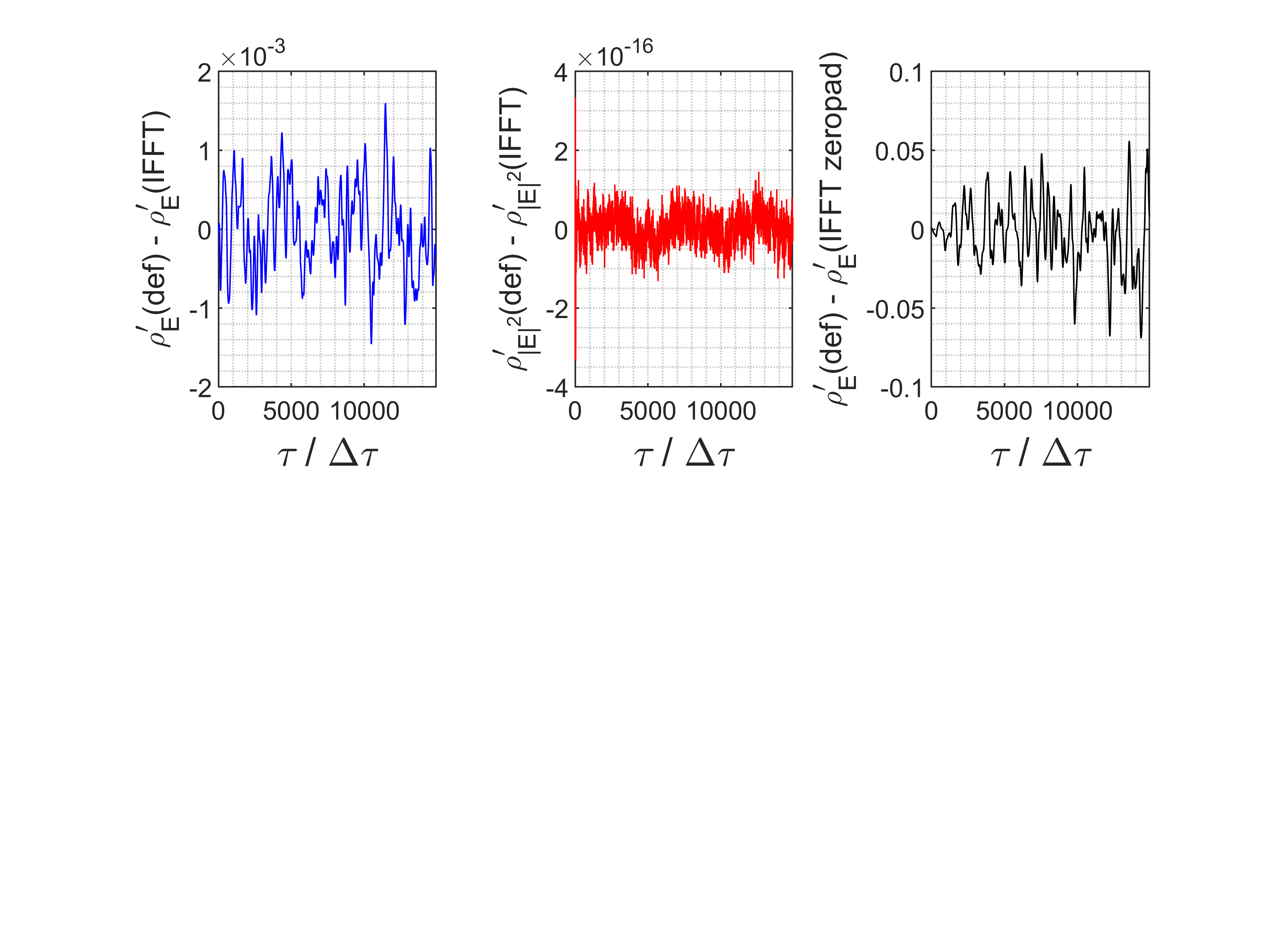}\\
\vspace{-4.5cm}\\
\\
~~~~~(i)~~~~~~~~~~~~~~~~~~(ii)~~~~~~~~~~~~~~~~~~(iii)\\
(b)\\
\end{tabular}
\end{center}
{
\caption{\label{fig:acf_def_vs_periodogram}
\small
{
(a) Definition-based (solid lines) vs. periodogram-based (symbols) ACFs of $E^{\prime(\prime)}$, $E$, $|E|^2$; 
(b) ACF residuals: (i)-(ii) between definition-based and IFFT-based circular ACFs, and (iii) between definition-based and zero-padded IFFT-based ACFs, at $f=2$ GHz.
}
}
}
\end{figure}

Fig. \ref{fig:acf_def_vs_periodogram}(a) compares various sample ACFs of $E^{(\prime(\prime))}(\tau)$ and $|E(\tau)|^2$ data at $f=2$ GHz using definition- vs. periodogram-based methods. The ACF of $E$ holds the middle between those of $E^\prime$ and $E^{\prime\prime}$, showing close correspondence with the IFFT of the periodogram for $E$ in Fig. \ref{fig:acf_def_vs_periodogram}(b)(i). Even closer agreement is found for the ACF of $|E|^2$ in Fig. \ref{fig:acf_def_vs_periodogram}(b)(ii). Fig. \ref{fig:acf_def_vs_periodogram}(a) also confirms the theoretical relationship $
\rho_{|E|^2} (\tau) \simeq \left [ \rho_{E^{\prime(\prime)}} (\tau) \right ]^2
$ valid for ideal circular fields.

An IFFT-based calculation produces the circular ACF, where data are presumed to originate from a periodically extended sequence (period $N_s$), as opposed to the linear ACF of an infinite aperiodic time series. This represents the physical reality in a mechanically stirred MSRC, where field values ideally repeat exactly after each full rotation of the stirrer. 
To remove the effect of ACF circularity, the data can be padded with $N_s$ zeros prior to performing the IFFT. Thus, taking measurement data across multiple rotations and comparing their linear ACF from its definition with the circular ACF of data for one rotation allows for detection of departures from ideal periodicity and slow\footnote{For investigation of nonstationarity within one period (fast drift or fast fading), local methods of analysis are required, e.g., using wavelets.} mechanical drift.

Fig. \ref{fig:acf_def_vs_periodogram}(b)(iii) shows the residuals between the definition- and zero-padded IFFT-based ACFs of $E$ at $f=2$ GHz, indicating that the maximum extent of correlation circularity is of the order of $5\%$, which decreases to $3\%$ at $f=18$ GHz. 
This provides a target figure for ACF approximation accuracy and indicates a $[1-(\Delta \tau/N_s)]^{-1}$ dependence, on average. 

\subsection{CCF}
For centered QC random fields with no deterministic component, $\rho^{\prime\prime}_E (\tau) \simeq [ \rho_{E^{\prime\prime},E^\prime} (\tau)- \rho_{E^\prime,E^{\prime\prime}} (\tau) ] / 2 \simeq \rho_{E^{\prime\prime},E^{\prime}} (\tau)$ vanishes at $\tau=0$, i.e., simultaneous  $E^\prime(\tau)$ and $E^{\prime\prime}(\tau)$ are uncorrelated. Since the CCF measures the linear cross-dependence as a function of $\tau$, it evaluates the phase shift between I- and field Q-components at different lags. This can be used for extracting an unstirred or line-of-sight component. A nonzero cross-correlation at some $\tau$ determines an average level of I/Q correlation, indicative of mean modulation. 

Another application is the detection of the stir period in a stir sweep that was extended beyond a single full turn of the stirrer: by cross-correlating sweep data in a window of sample length $N$ with data in a stepwise sliding window, this yields the start of the next period at the location $\tau_s$ where the cross-correlation shows a maximum (peak) value. This technique is robust to small levels of noise and drift.

\begin{figure}[htb] 
\begin{center}
\begin{tabular}{c}
\vspace{-0.5cm}\\
\hspace{-0.6cm}
\includegraphics[scale=0.65]{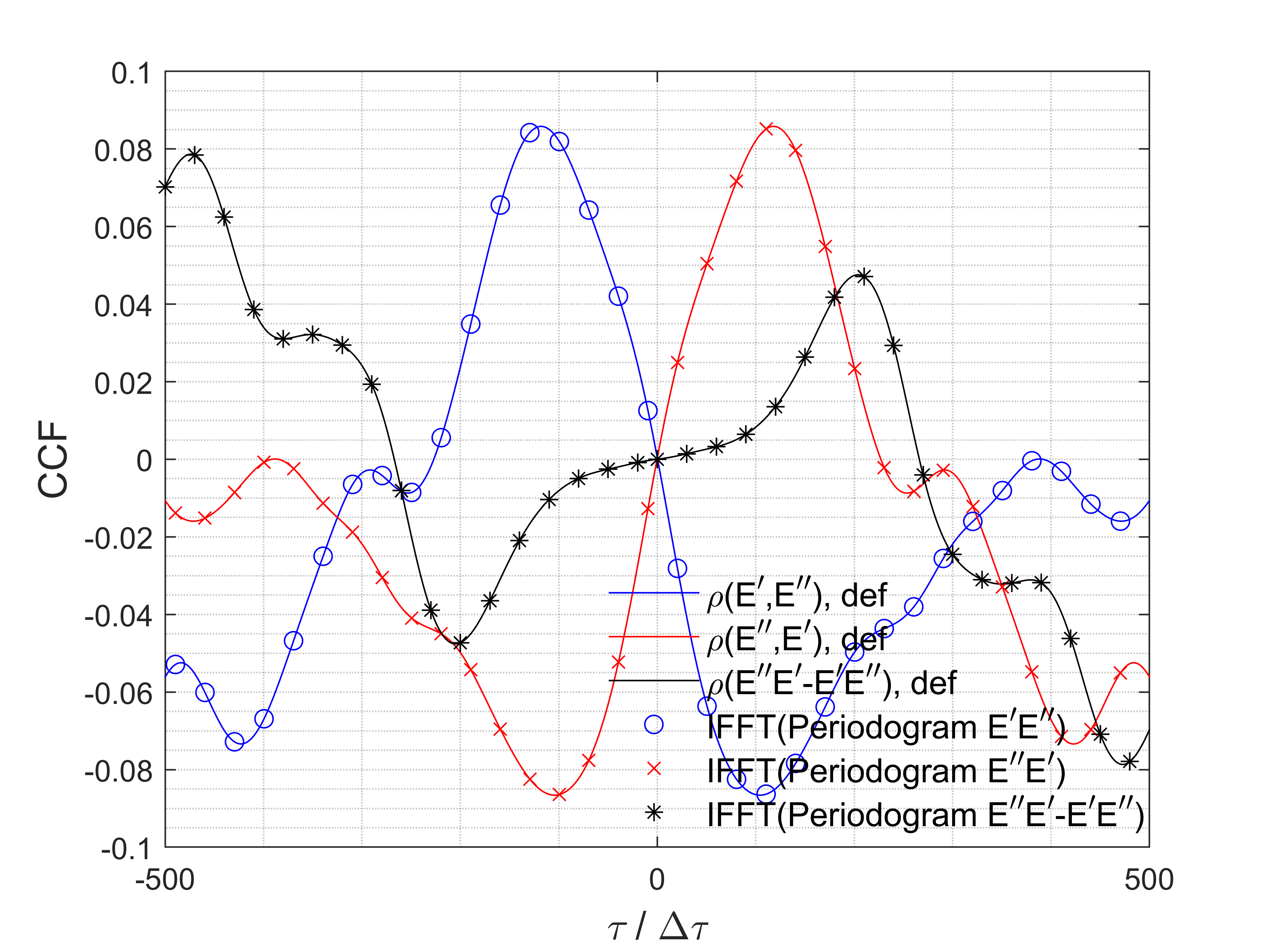}\\
\vspace{-1cm}\\
\\
(a)\\
\hspace{-0.8cm}
\includegraphics[scale=0.65]{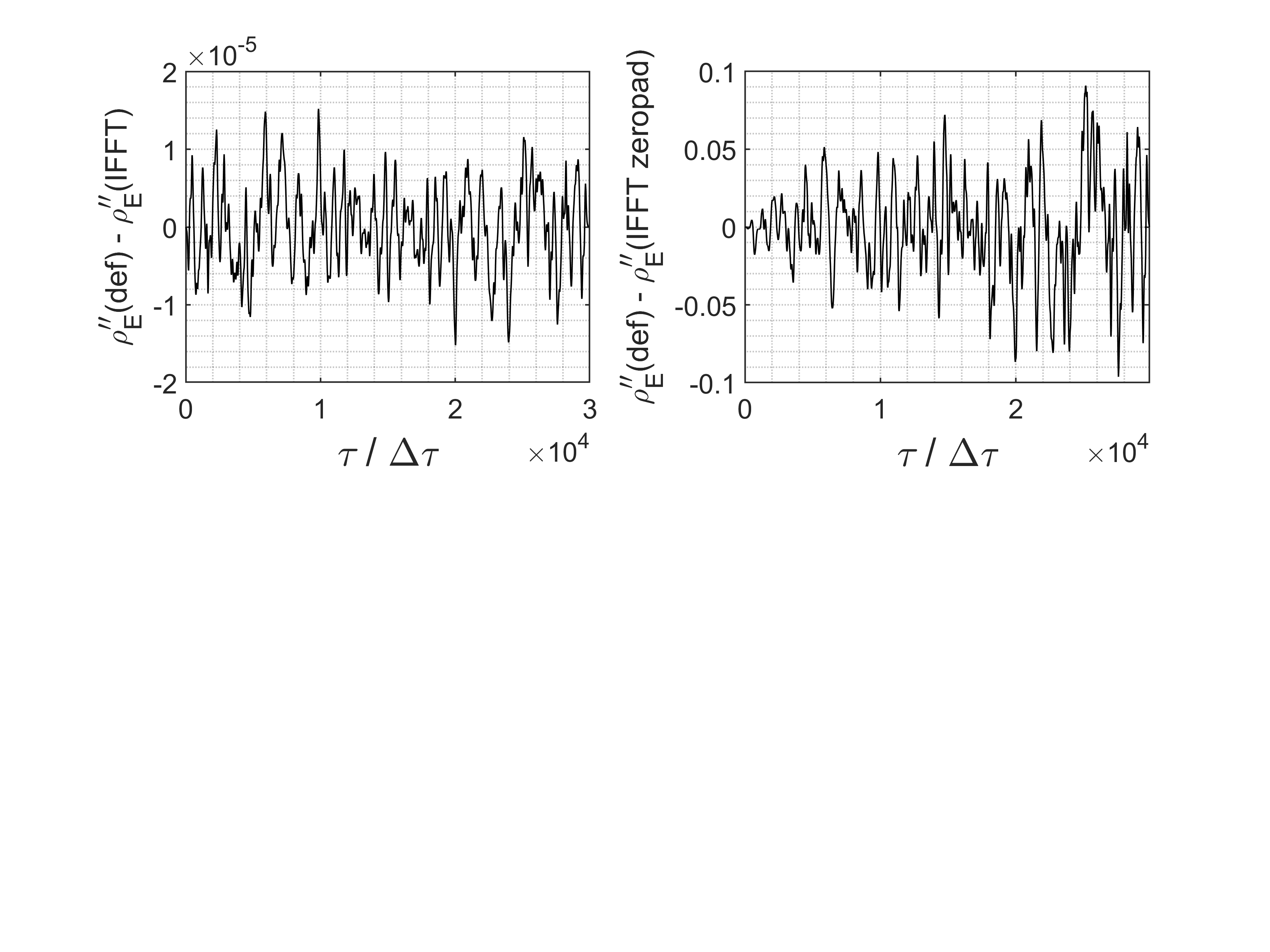}\\
\vspace{-4.5cm}\\
\\
~~~~~(i)~~~~~~~~~~~~~~~~~~~~~~~~~~~~(ii)\\
(b)\\
\end{tabular}
\end{center}
{
\caption{\label{fig:ccf_def_vs_periodogram}
\small
{
(a) Definition-based (solid lines) vs. periodogram-based (symbols) CCFs of $E$; 
(b) CCF residuals: (i) between definition-based and IFFT-based circular CCFs, and (ii) between definition-based and zero-padded IFFT-based CCFs, at $f=2$ GHz.
}
}
}
\end{figure}

\begin{figure}[htb] 
\begin{center}
\begin{tabular}{c}
\hspace{-0.7cm}
\includegraphics[scale=0.67]{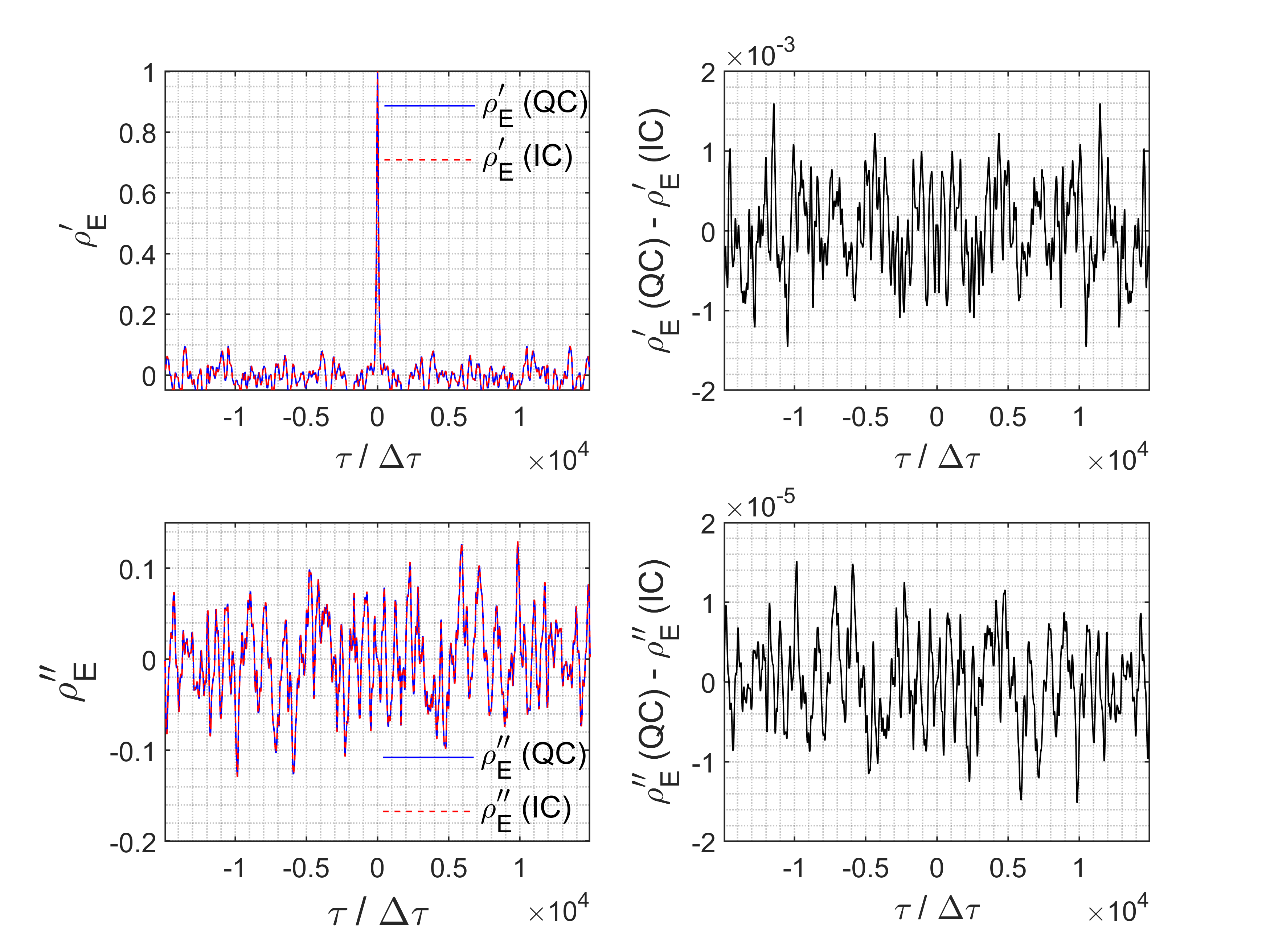}\\
\end{tabular}
\end{center}
{
\caption{\label{fig:acf_ccf}
\small
{
ACF (top left) and CCF (bottom left) of $E(\tau)$ at $f=2$ GHz, tune state 1 of 72: ideal IC vs. actual QC approximation, with residuals of ACF (top right) and CCF (bottom right).
}
}
}
\end{figure}

Fig. \ref{fig:ccf_def_vs_periodogram} shows measurement results of CCFs at $f=2$ GHz, complementing those for the ACF. 
Fig. \ref{fig:acf_ccf} compares ACFs and CCFs of $E$ for assumed ideal circular (IC) fields, i.e.,
\begin{align}
\rho^{\rm (IC)}_E(\tau) = \frac{\sigma_{E^\prime,E^\prime}(0,\tau) + \sigma_{E^{\prime\prime},E^{\prime\prime}}(0,\tau)}{\sigma^2_{E^\prime}(0) + \sigma^2
_{E^{\prime\prime}}(\tau)}
\label{eq:CCF_exact}
\end{align}
with the approximation of (\ref{eq:CCF_exact}) for QC uniform fields
\begin{align}
\rho^{\rm (QC)}_E(\tau) &\simeq \frac{1}{2} 
\left \{ 
\left [ 
\rho^\prime_{E^{\prime      },E^{\prime      }}(0,\tau) + 
\rho^\prime_{E^{\prime\prime},E^{\prime\prime}}(0,\tau) \right ] \right . \nonumber\\
&~~~~~~~ \left . 
+ \rmj 
\left [ 
\rho^{\prime\prime}_{E^{\prime\prime},E^{\prime}}(0,\tau)
- 
\rho^{\prime\prime}_{E^{\prime},E^{\prime\prime}}(0,\tau) 
\right ] 
\right \}
.
\label{eq:CCF_approx}
\end{align}
The plots indicate that IC is closely approximated for the present data, with deviations being of the order of $1\%$ for $\rho^\prime_E(\tau)$ and $0.01\%$ for $\rho^{\prime\prime}_E(\tau)$. 

\subsection{Pad\'{e} vs. Taylor Approximation Models for CFs}
\begin{figure}[htb] 
\begin{center}
\begin{tabular}{c}
\hspace{-0.6cm}
\includegraphics[scale=0.65]{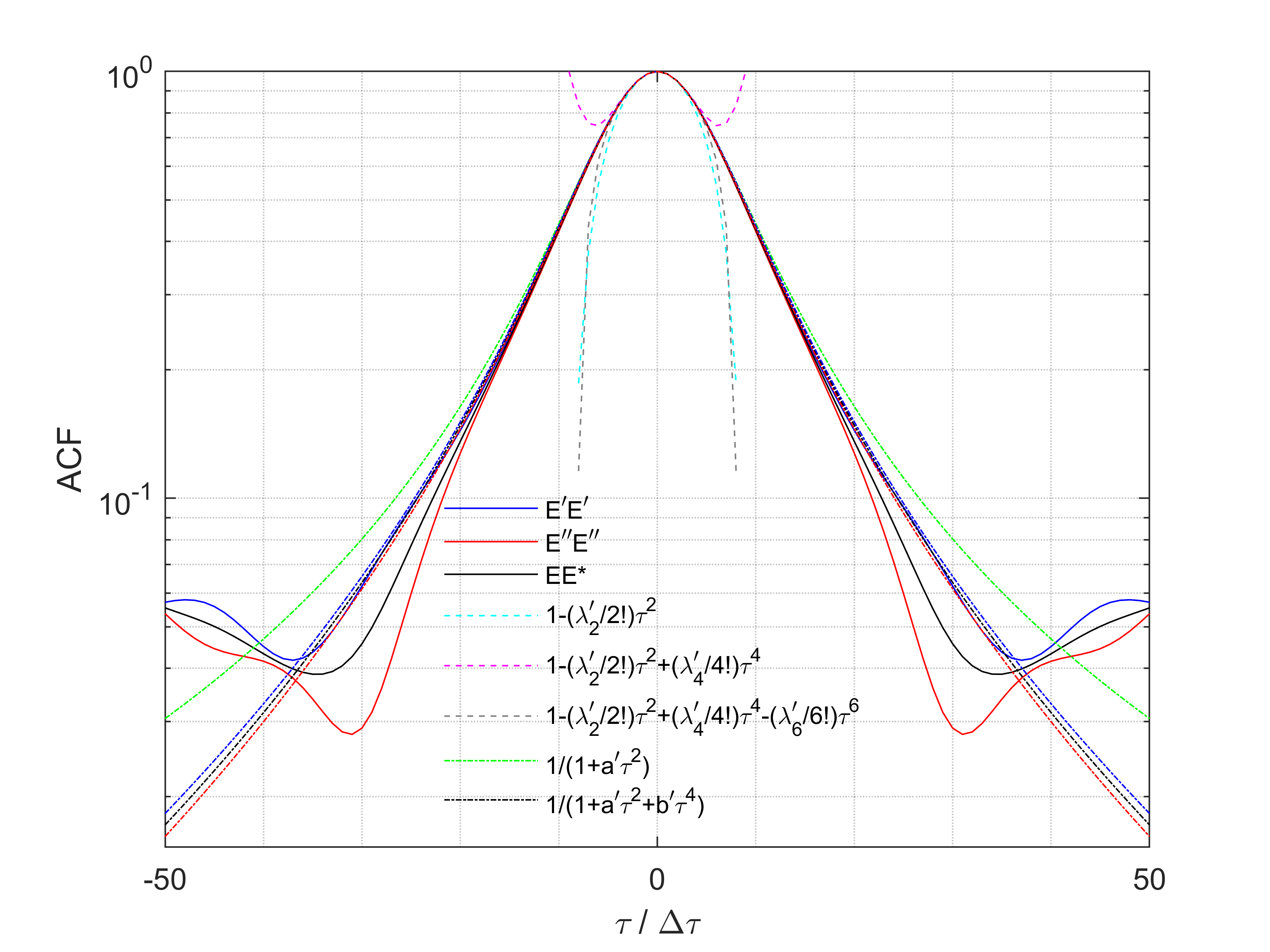}
\\
\vspace{-0.5cm}\\
(a)\\
\hspace{-0.6cm}
\includegraphics[scale=0.65]{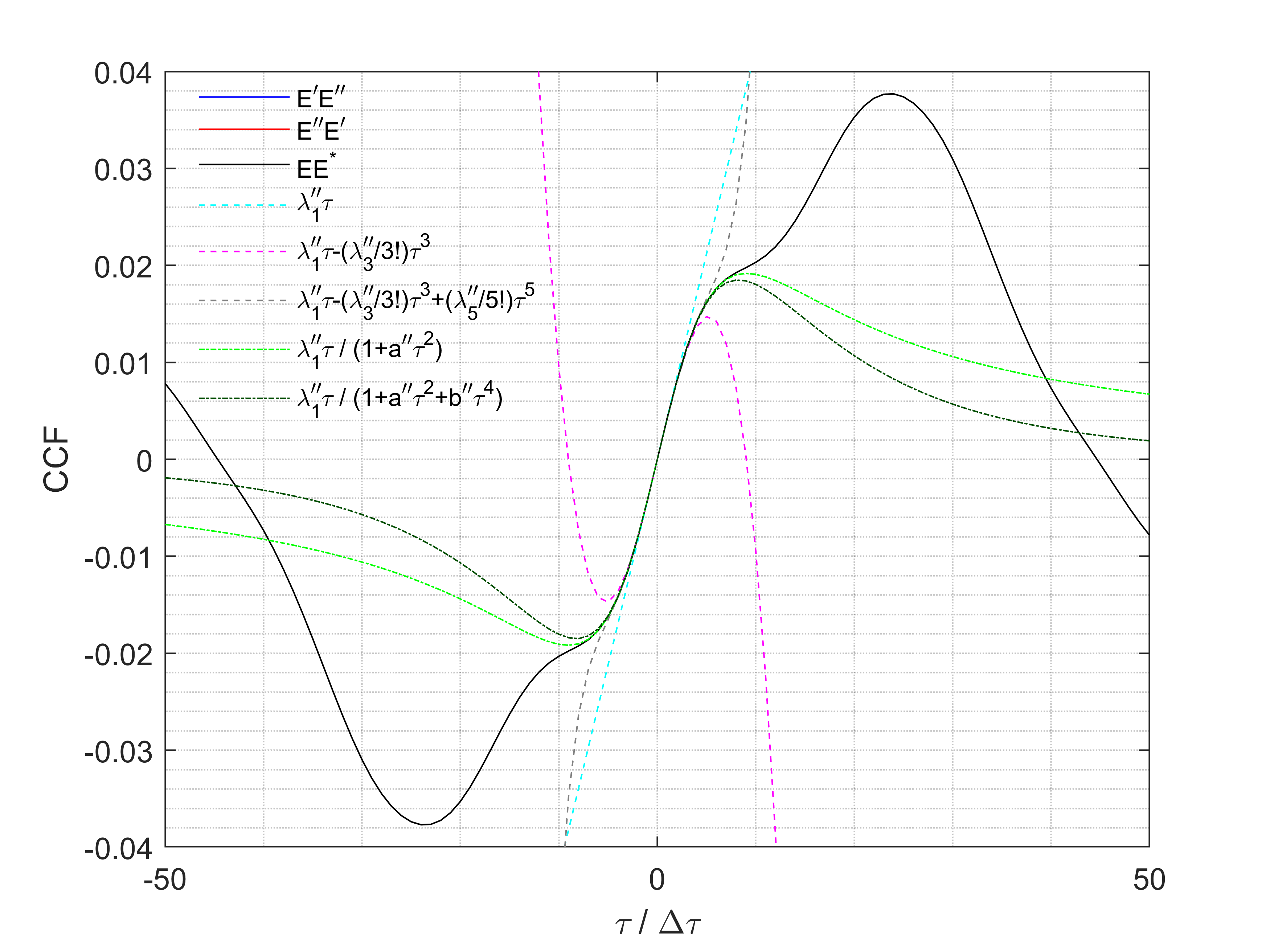}\\
\vspace{-0.5cm}\\
(b)
\end{tabular}
\end{center}
{
\caption{\label{fig:acf_ccf_Taylor_vs_Pade}
\small
{
(a) I/I, Q/Q measured ACFs and (b) I/Q, Q/I measured CCFs of $E^{\prime(\prime)}$ and $E$ (solid), compared with 1st-, 2nd- and 3rd-order models for Taylor (dashed), 1st-order Pad\'{e} (green dot-dashed for $E$) and 2nd-order Pad\'{e} (black dot-dashed for $E$; other dot-dashed with same color coding as measured ACFs for $E^{\prime(\prime)}$), at $f=18$ GHz.
}
}
}
\end{figure}

Fig. \ref{fig:acf_ccf_Taylor_vs_Pade}(a) compares first- and higher-order Taylor and Pad\'{e} approximations of $\rho^\prime_E(\tau)$ \cite[eqs. (22) and (25)]{arnaACFSDF_pt1} to the actual (i.e., sample) ACF at $f=18$ GHz. The ACF $\rho^\prime_{E,E^*}(\tau)$ lies midway between $\rho_{E^\prime,E^\prime}(\tau)$ and $\rho_{E^{\prime\prime},E^{\prime\prime}}(\tau)$, as expected. The $[0/2]$-order (in $\tau^2$) Pad\'{e} approximation closely follows the experimental sample ACF over a broad $\tau$-domain, significantly outperforming the Taylor expansions in $\tau^2$. Even a mere $[0/1]$-order approximation offers a superior match across a broader domain than a third-order Taylor expansion. 
Specifically, the correlation length associated with the 1/e-level of the actual ACF is approximately $\tau/\Delta\tau \simeq 10$, serving as a reference value. In comparison, second- and third-order Taylor approximations exhibit a close match up to three lag units only and correlation level $0.7$. By contrast, first- and second-order Pad\'{e} approximants offer a close match up to $15$ and $25$ lag units and down to levels as low as $0.2$ and $0.1$, respectively. 
Even larger $\tau$-domains for close correspondence occur at lower frequencies, but are also found for less efficient stirring.

Fig. \ref{fig:acf_ccf_Taylor_vs_Pade}(b) shows first- and higher-order approximations to the actual sample CCF. 
The deviation of $\rho^{\prime\prime}_E(\tau/\Delta\tau)$ from linearity sets in after just a few units of $\tau/\Delta\tau$, indicating nonlinearity of the stir process from this point onward.
Similar to the ACFs, the Pad\'{e} approximations again offer an improved CCF fit compared to Taylor expansions, although now more rapidly diverging outside the region of convergence (cross-correlation length $\tau^{\prime\prime}_c$), yet still offering an improved fit, even for a reduced Pad\'{e} order. 
Importantly, and unlike Taylor expansions of any order, the Pad\'{e} approximants also exhibit a correct asymptotic dependence for $|\tau|\rightarrow  +\infty$ in both the ACF and CCF.

In summary, the $\tau$-domain where a good fit for CFs can be achieved is considerably larger for a Pad\'{e} than for a Taylor approximation, for equal orders of approximation and beyond. Inevitably, this domain remains restricted because both types of expansions are point-wise around $\tau=0$. For the purpose of approximating $\rho^{\prime(\prime)}_E(|\tau|  \leq \tau^{\prime(\prime)}_c)$ in EMC/EMI applications \cite{arnalocavg}, \cite{arnathresh}, this is sufficient because the ACF level at $\tau^\prime_c$ is well below 1/e, even for the simplest $[0/1]$ model. Beyond this domain, other methods for modelling CFs can be adopted, e.g., using interpolating functions.

\section{Field Spectral Densities}
\subsection{ASDF\label{sec:ASDF}}
\subsubsection{Spectrum}
\begin{figure}[htb] 
\begin{center}
\begin{tabular}{c}
\hspace{-0.6cm}
\includegraphics[scale=0.65]{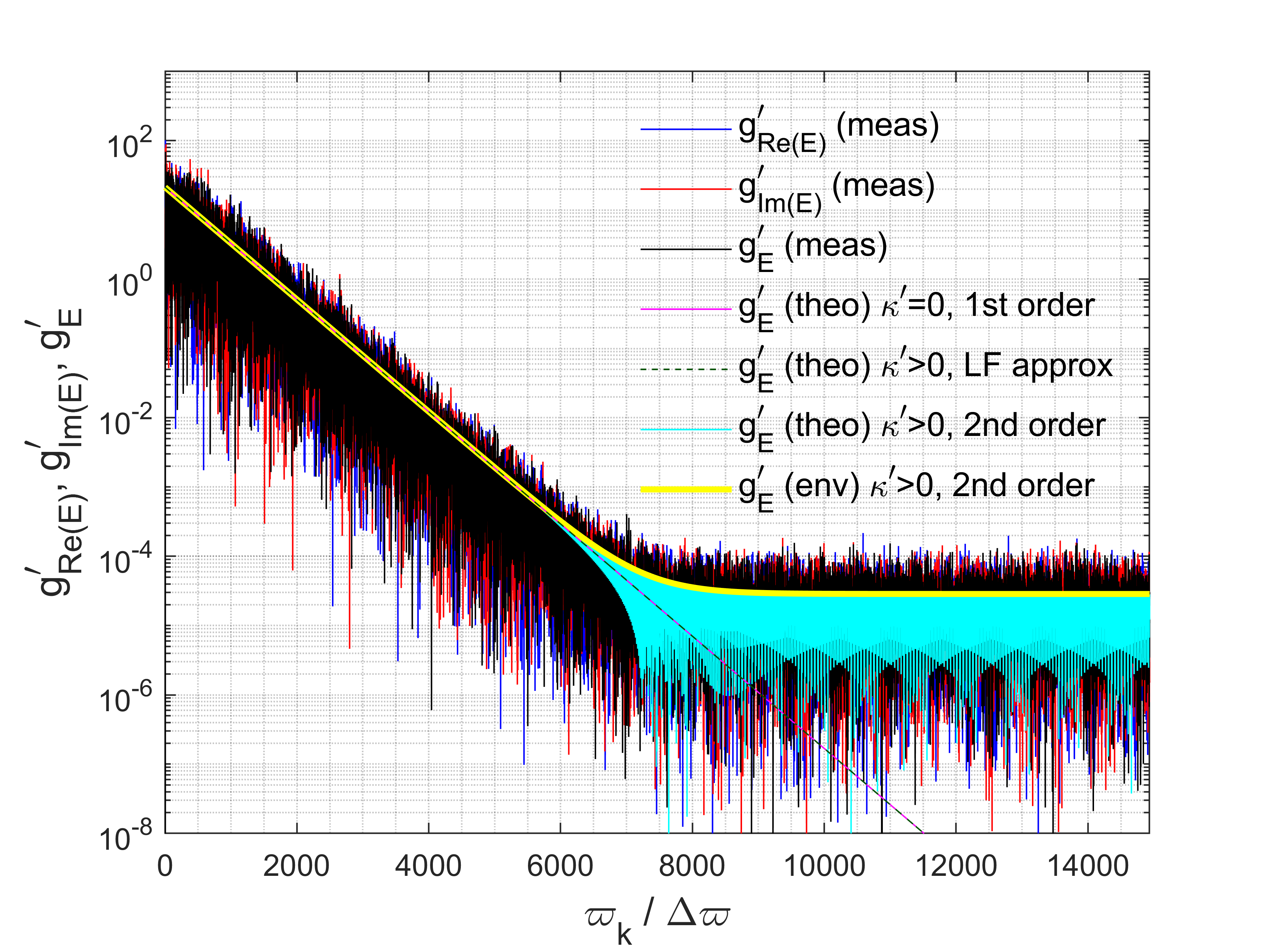}\\
\end{tabular}
\end{center}
{
\caption{\label{fig:fsdf_WK_vs_periodogram}
\small
{
Comparison of experimental vs. theoretical 
ASDFs $g^\prime_E(\varpi_k/\Delta\varpi)$: 
1st-order exponential ($\kappa^\prime=0$), LF-approximating, and
2nd-order Pad\'{e}-based models ($\kappa^\prime = 5\times 10^{-13}$), and envelope of the ASDF
as a function of $k\equiv\varpi_k/\Delta \varpi$ for $0 \leq k \leq \lfloor N_s/2 \rfloor$ and $\Delta \varpi = 2\pi/(N_s\Delta\tau)$, at $f=18$ GHz.
}
}
}
\end{figure}

Since no spectral windowing is applied, the ASDF coincides with the periodogram.
Fig. \ref{fig:fsdf_WK_vs_periodogram} compares the empirical ASDF $g^\prime_E(\varpi)$ for the measured $E(\tau)$ at $f=18$ GHz, obtained by starting from the ACF and application of the EWK theorem \cite{arnaACFSDF_pt1} or the periodogram (\ref{eq:periodgram_p}), vs. first- and second-order Pad\'{e}-based ASDF models and envelope \cite[eqs. (34), (38), (39)]{arnaACFSDF_pt1}, showing good agreement. 
The simple first-order model provides an estimate of the mean level and rate of low-frequency (LF) decay; the second-order model additionally offers an estimate for the stir high-frequency (HF) asymptotic level and the DC-to-Nyquist level drop of the ASDF. 

\begin{figure}[htb] 
\begin{center}
\begin{tabular}{c}
\hspace{-0.5cm}
\includegraphics[scale=0.65]{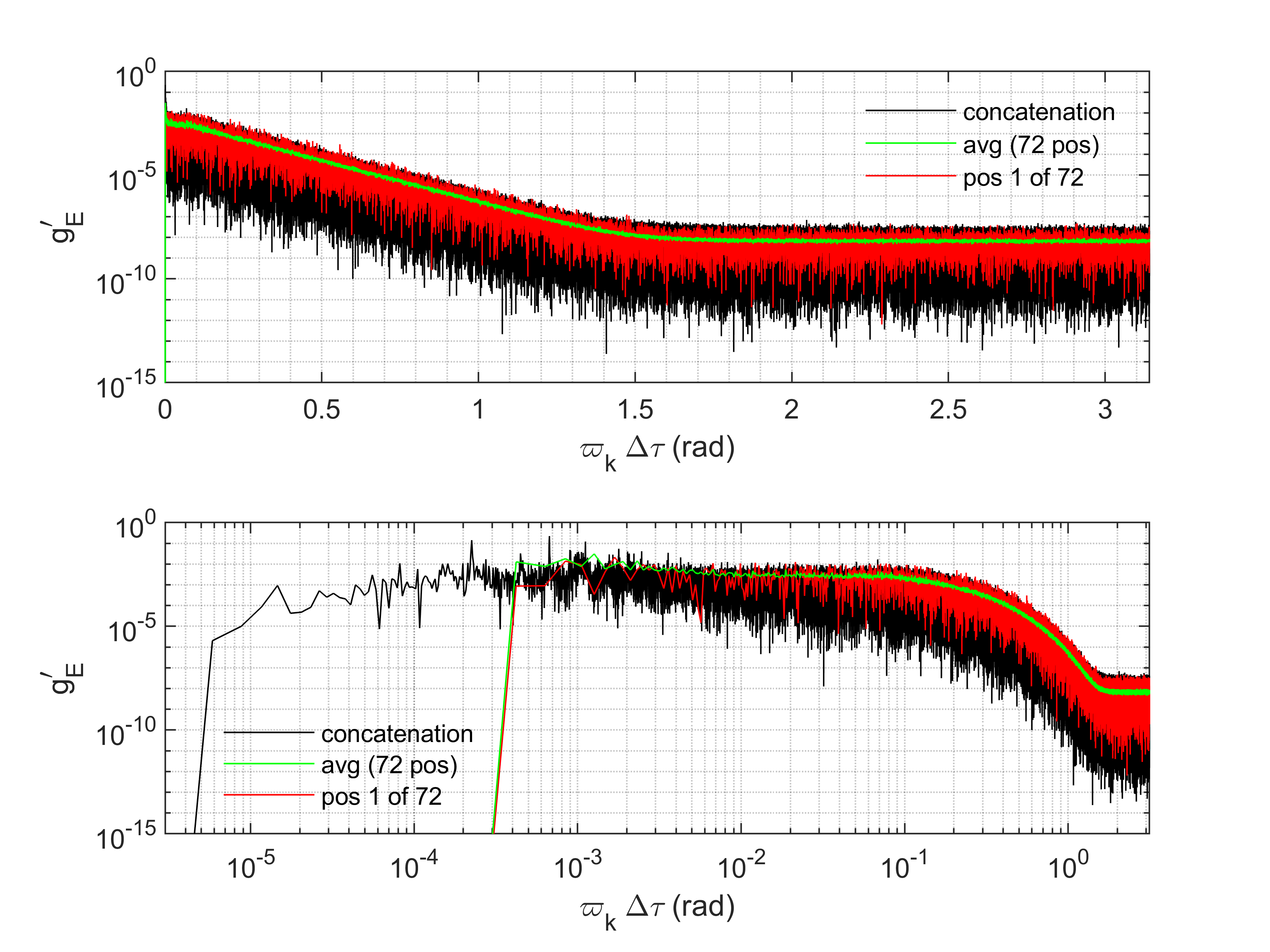}
\end{tabular}
\end{center}
{
\caption{\label{fig:ASDF_SpinLarge_concat_f18GHz}
\small
{
Periodogram-based field ASDFs $g^\prime_E(\varpi_k \Delta\tau)$ for stir sweeps by large paddle and tune states of small paddle: 
(i) single sweep at single tune state (1 of 72); 
(ii) averaged ASDFs across 72 tune states; 
(iii) concatenation of 72 stir sweeps, where $\varpi_{k} =  k\Delta\varpi = 2k\pi/(N\Delta\tau)$, $0 \leq k \leq \lfloor N/2 \rfloor$, $\varpi_{\rm max} = \varpi_{N/2} = \pi/\Delta\tau$ rad/s.
Top: linear frequency scale, bottom: logarithmic frequency scale.
}
}
}
\end{figure}

Fig. \ref{fig:ASDF_SpinLarge_concat_f18GHz} compares normalized ASDFs for stirring by the large paddle, obtained as periodograms in three different ways: 
(i) as a sample ASDF, at an arbitrary angular tune state of the secondary small paddle (tuner step size $\Delta\tau_2 = 5$ deg), from a single stir sweep yielding $N=N_s = 29\,869$ stir frequencies; 
(ii) as the ensemble averaged ASDF $\langle g^\prime_E(\varpi_k \Delta\tau)\rangle \equiv \langle g^\prime_E(\varpi_k \Delta\tau)\rangle_{N_t}$, from all $N_t=72$ individual ASDFs, averaged per $\varpi_k$ ($N=N_s = 29\,869$ stir frequencies);
(iii) from the ``super'' stir sweep constructed by concatenating all 72 tuner sweeps ($N=N_t N_s = 72 \times 29\,869$).
For (ii) and (iii), each $i$th sweep is weighted by its own variance $\lambda_{0,i}$ prior to aggregation ($i=0,\ldots,71$).
Concatenation (or, alternatively, increasing $N_s$ for a single stir sweep) increases spectral resolution within the ASDF, but does not decrease the mean square fluctuation across $\varpi_k$, even though it decreases the covariances $\langle \cE^{\prime(\prime)}(\varpi_k) \cE^{\prime(\prime)}(\varpi_\ell) \rangle$, on average \cite{prie1981}. 

The value of  $\langle g^\prime_E(\varpi_k \Delta\tau)\rangle$ at stir DC is zero because, in this processing and unlike in \cite{arnalocavg}, field data were centered by removal of the stir averaged (i.e., unstirred) field {\it a priori}. 
In Fig. \ref{fig:ASDF_SpinLarge_concat_f18GHz}, $\langle g^\prime_E(\varpi\Delta\tau)\rangle $ reaches its maximum just in excess of $\varpi \Delta \tau \simeq 0.001$ rad, then decays slowly until $\varpi \Delta \tau \simeq 0.01$ rad, stays approximately constant up to $\varpi \Delta \tau \simeq \pi/40$ rad, before decaying exponentially until $\varpi^\prime_c \Delta \tau \simeq \pi/2$ rad. 
The flattening of $\langle g^\prime_E(\varpi\Delta\tau)\rangle $ at LFs is caused in part by the finite length of the time window (stir period). 
For $0.01 \leq \varpi \Delta \tau \simeq \pi/40$ rad, the slope of $\langle g^\prime_E(\varpi_k)\rangle$ approaches zero quadratically.
Since a leading-order quadratic dependence of $\langle g^\prime_E(\varpi)\rangle$ on $\varpi$ near $\varpi=0$ implies a finite second-order autocorrelation moment
\begin{align}
\theta^\prime_2 \stackrel{\Delta}{=} \int^{+\infty}_{-\infty} \tau^2 \rho^\prime_E(\tau) \rmd\tau < +\infty
\end{align} 
this suggests that higher-order Pad\'{e} approximants for the ACF may offer further model refinement for the exponential decay of the ASDF near $\varpi=0$ (faster ACF decay for $|\tau/\tau^\prime_\rmc|\gg 1$).

In \cite{arnaACFSDF_pt1}, theoretical HF asymptotic values for continuous ASDFs for the $[0/2]$ Pad\'{e} approximant were derived. For discrete ASDFs, Tbl. \ref{tbl:asymp_comparison} compares corresponding values at the Nyquist rate with those extracted from the empirical ASDF. As expected, the DC-to-Nyquist level drop is highly sensitive to the value of $\kappa^\prime$, whose value is difficult to estimate accurately from sampled data, even after compensation for finite-difference bias; cf., \cite{arnaACFSDF_pt2}. If, instead, a smaller positive value of $\kappa^\prime = 5 \times 10^{-13}$ is assigned and $\lambda^\prime_4$ is deduced accordingly -- in this case increased by merely $0.65\%$ -- while maintaining $\lambda^\prime_2$, closer agreement can be found, as shown in Tbl. \ref{tbl:asymp_comparison}. Inclusion of further higher-order Pad\'{e} approximant terms and/or increase of $f$ leads to further improved agreement.

\def\arraystretch{1.3}
\begin{table}
\begin{center}
\begin{tabular}{||l||l|l|l||}\hline\hline
Parameter & [0/2] Pad\'{e} & [0/2] Pad\'{e}, & measured \\ 
~ & ~ & set value $\kappa^\prime$ & ~ \\ \hline\hline
$(\lambda^{\prime}_2)^{1/2}$ (rad/s) & $595.6$   & $595.6$ & \\ 
$(\lambda^{\prime}_4)^{1/4}$ (rad/s) & $926.3$   & $932.1644$ & \\ 
FD debiased $\kappa^\prime$              & $0.01118$ & $5{\times}10^{-13}$ & \\ \hline \hline
$g^\prime_E(\pi)/g^\prime_E(0+)$ (dB)     & $-9.7$ & $-61.5$ & $-60.1$ \\ \hline
$\varpi^\prime_c \Delta\tau$ (rad)  & $1.22$ & $1.25$ & $1.51$ \\ \hline
$\dot{g}^\prime_E(\varpi \ll \varpi^\prime_c)$ (dB/rad) & $-52.3$ & $-52.5$ & $-43.0$ \\ \hline \hline
\end{tabular}
\caption{\label{tbl:asymp_comparison} \small 
DC-to-Nyquist level drop $g^\prime_E(\pi)/g^\prime_E(0+)$, normalized corner frequency $\varpi^\prime_c \Delta\tau$, and LF slope $\dot{g}^\prime_E(\varpi \ll \varpi^\prime_c)$ for $g^\prime_E(\varpi)$ averaged across 72 tune states, at $f=18$ GHz: measured results vs. [0/2] Pad\'{e}-based ASDF results for finite-difference (FD) debiased value $\kappa^\prime=0.01118$ vs. 
inferred value $\kappa^\prime\stackrel{\Delta}{=}5 \times 10^{-13}$.
}
\end{center}
\end{table}

\subsubsection{Stir Spectral Coefficient of Variation}
The ensemble of stir SDFs generated by varying the tune states of the secondary paddle, $\tau_2$, allows for constructing a 2-D stir spectrogram; cf. sec. \ref{sec:interchange}.
For statistically equivalent secondary tune states, simpler 1-D summary representations are provided by summary statistics of the SDF, e.g., $\langle g^\prime_E(\varpi_k)\rangle$ and $\sigma_{g^\prime_E}(\varpi_k)$ from aggregation across all tune states, at arbitrary $\varpi_k$. 

\begin{figure}[htb] \begin{center}
\begin{tabular}{c}
\hspace{-0.6cm}
\includegraphics[scale=0.65]{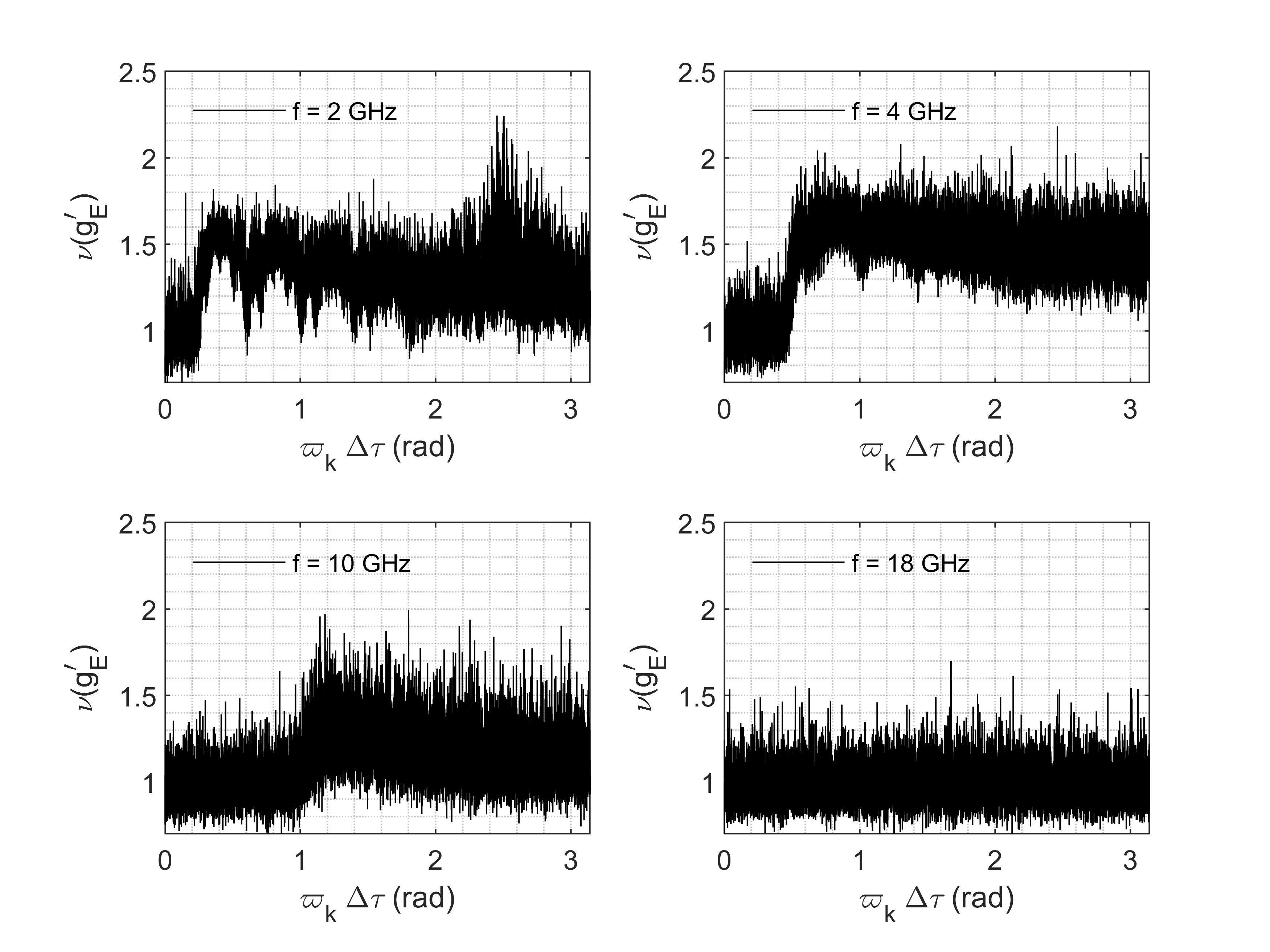}\\
\end{tabular}
\end{center}
{
\caption{\label{fig:CffVarSDF_SpinLarge_f18GHz}
\small
{Stir spectral coefficient of variation $\nu_{g^\prime_E} (\varpi_k \Delta\tau)$ for $\langle g^\prime_E(\varpi_k\Delta\tau; \tau_{2,\ell})\rangle_{N_t}$, $0 \leq k \leq \lfloor N_s /2 \rfloor$ at $f=2,4,10,18$ GHz.
}
}
}
\end{figure}

Fig. \ref{fig:CffVarSDF_SpinLarge_f18GHz} shows the coefficient of variation, $\nu_{g^\prime_E}(\varpi_k \Delta\tau) \stackrel{\Delta}{=} \sigma_{g^\prime_E}(\varpi_k \Delta\tau) / \langle g^\prime_E(\varpi_k \Delta\tau) \rangle$. At $f=18$ GHz across all $\varpi_k \Delta\tau > 0$, its values fluctuate uniformly around 1, corresponding to $\chi^2_2$-distributed $|\cE(\varpi_k>0;\tau_2)|^2$, whereas $|\cE(\varpi_k=0;\tau_2)|^2$ is $\chi^2_1$-distributed \cite{jenk1968}. 
The 72 tune samples $\cE(\varpi_k;\tau_{2,\ell})$ for arbitrary $\varpi_k$ form a sample set from a (quasi-)circular Gaussian distribution. 
At lower $f$, the values of $\nu_{g^\prime_E}(\varpi_k \Delta\tau)$ remain centered around 1 only for sufficiently low but positive $\varpi_k$, i.e., for large but finite sample separations in the stir domain. 
From some $f$-dependent transition stir frequency $\varpi_o$ upward, $\nu_{g^\prime_E}(\varpi_k \Delta\tau)$ increases sharply because the corresponding stir arcs are too small to provide good stir statistics. This $\varpi_o$ decreases with decreasing $f$, as the stir arc traversed during a given interval of $\tau$ becomes progressively shorter relative to the CW wavelength. 
For example, at $f=4$ GHz the jump in $\nu_{g^\prime_E}(\varpi_k \Delta\tau)$ occurs at $\varpi_o \Delta\tau\simeq 
\pi/6$ rad ($\sim\varpi_o/\Delta\varpi \simeq 2490$).

\subsection{CSDF}
\begin{figure}[htb] 
\begin{center}
\begin{tabular}{c}
\vspace{-0.5cm}\\
\hspace{-0.4cm}
\includegraphics[scale=0.65]{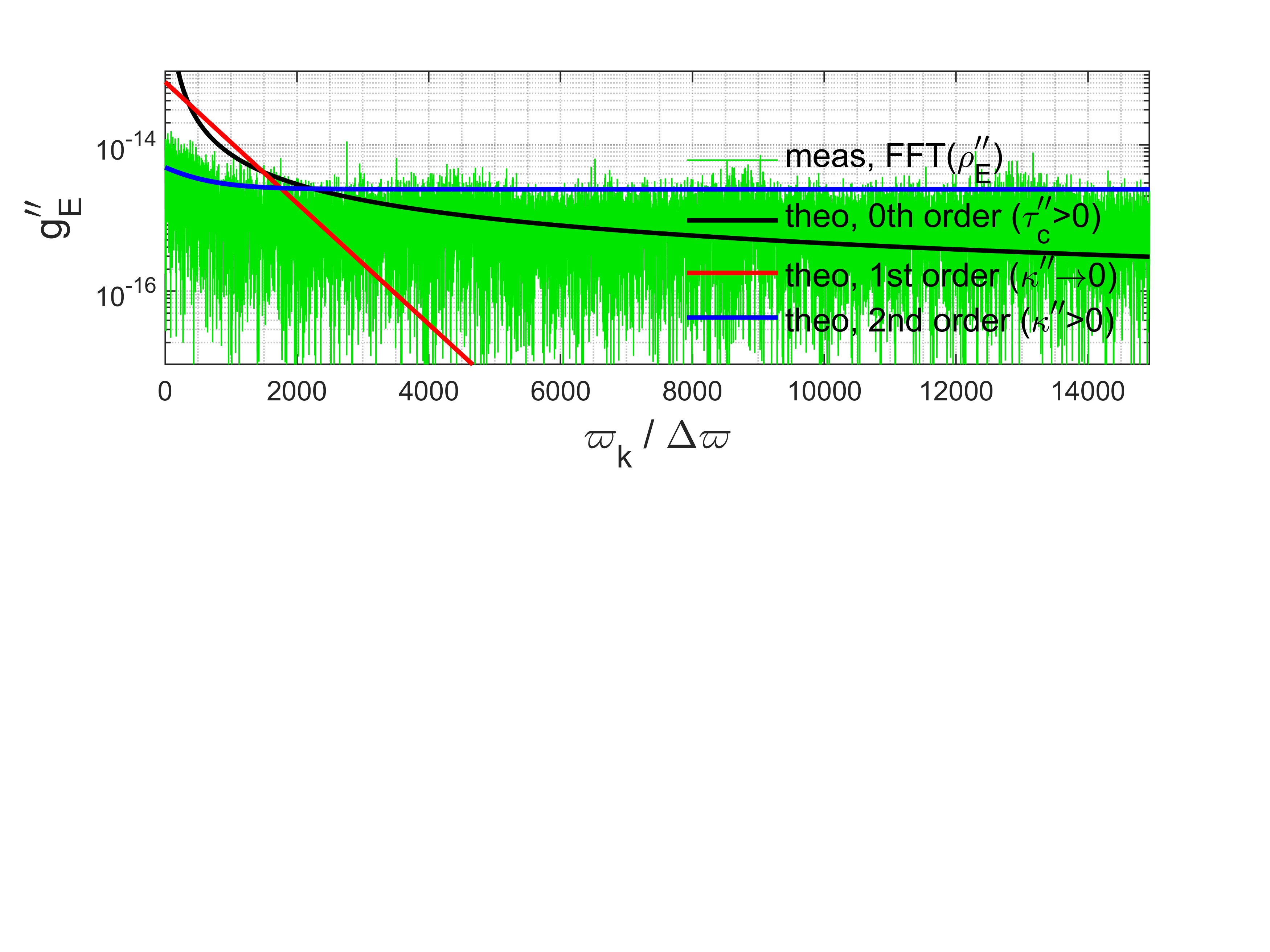}\\
\vspace{-4.8cm}\\
\\
\end{tabular}
\end{center}
{
\caption{\label{fig:crossfsdf_WK_vs_periodogram}
\small
{
Measured CSDF (EWK method) vs. theoretical 0th-, 1st-, and 2nd-order Pad\'{e}-based CSDF envelopes of $g^{\prime\prime}_E(\varpi_k/\Delta\varpi)$ for $\tau^{\prime\prime}_c = 5\Delta\tau = 1.328$ ms, ${\kappa^{\prime\prime}} \rightarrow 0$, and ${\kappa^{\prime\prime}} = 0.0854$, respectively, for stirring by large paddle at $f=18$ GHz. 
}
}
}
\end{figure}

The CSDF $g^{\prime\prime}_E (\varpi)$ represents the spectral  distribution of the I/Q covariance $\sigma_{E^\prime,E^{\prime\prime}}$. 
Fig. \ref{fig:crossfsdf_WK_vs_periodogram} compares envelopes of zeroth-, first-, and second-order Pad\'{e}-based CSDF models for $g^{\prime\prime}_E(\varpi_k /\Delta\varpi)$ against EWK-based\footnote{CSDFs obtained using the I/Q periodograms (not shown) were found to be nearly identical.} empirical CSDFs obtained by FFT of the experimental CCF, at $f=18$ GHz. 
The theoretical envelopes are derived from \cite[eqs. (57), (59) and (66)]{arnaACFSDF_pt1} and estimated debiased $\lambda^{\prime\prime}_i$ \cite{arnaACFSDF_pt2}. 
Increasing the Pad\'{e} order improves the agreement of LF and HF asymptotic regimes of the CSDF, on average. Further improvement for the LF part of the ASDF is also expected by futher increasing the order.

Comparing Fig. \ref{fig:crossfsdf_WK_vs_periodogram} for $g^{\prime\prime}_E(\varpi_k/\Delta\varpi)$ with Fig. \ref{fig:fsdf_WK_vs_periodogram} for $g^{\prime}_E(\varpi_k/\Delta\varpi)$ shows that the latter achieves a better fit. This is commensurate with general properties of discrete Fourier sine and cosine transformations, respectively, as the cosine transform typically offers a more efficient and accurate spectral representation, for arbitrary orders, compared to the former.

\begin{figure}[htb] 
\begin{center}
\begin{tabular}{c}
\hspace{-0.6cm}
\includegraphics[scale=0.65]{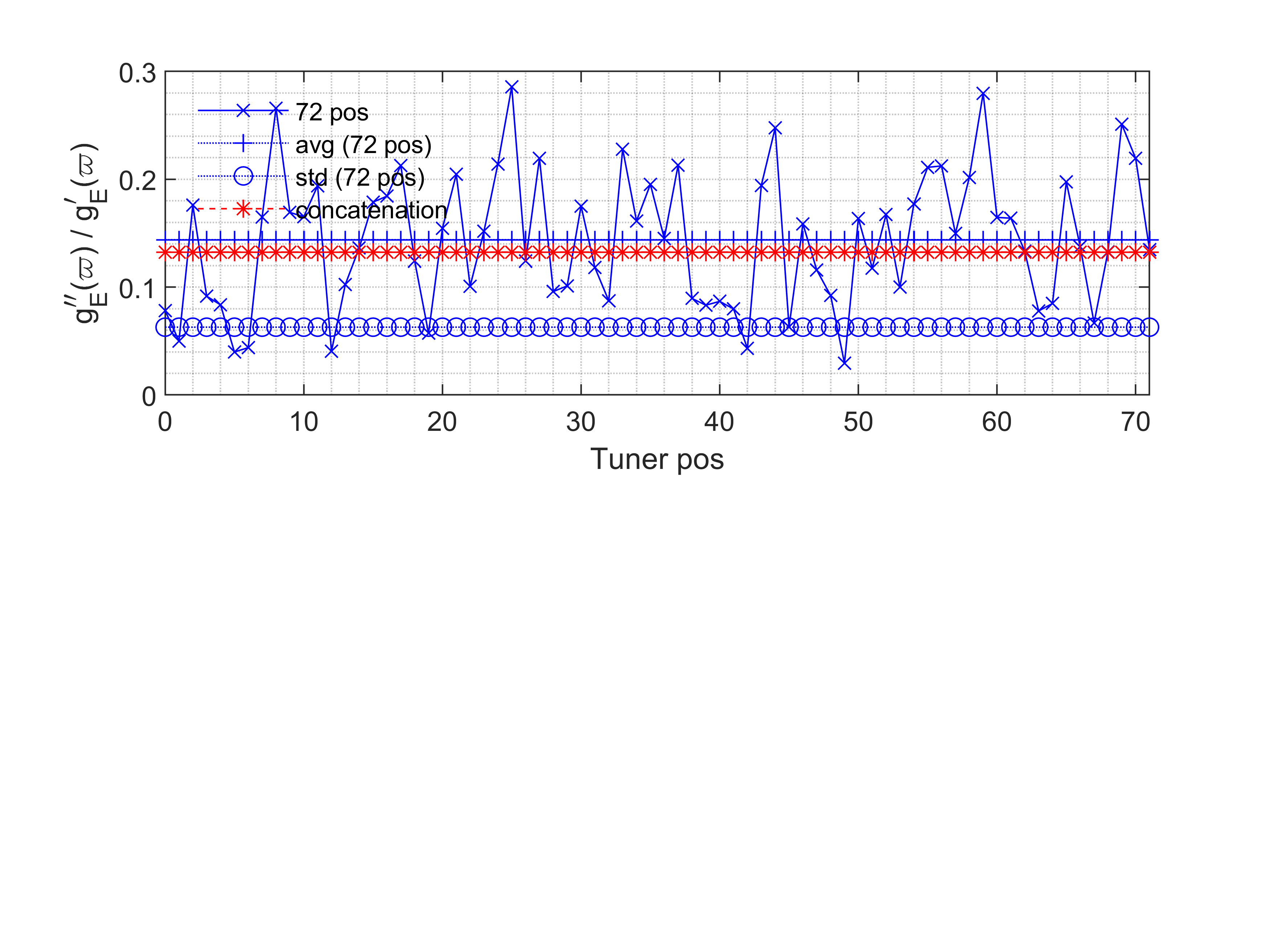}\\
\vspace{-4.3cm}\\
(a)\\
\hspace{-0.6cm}
\includegraphics[scale=0.65]{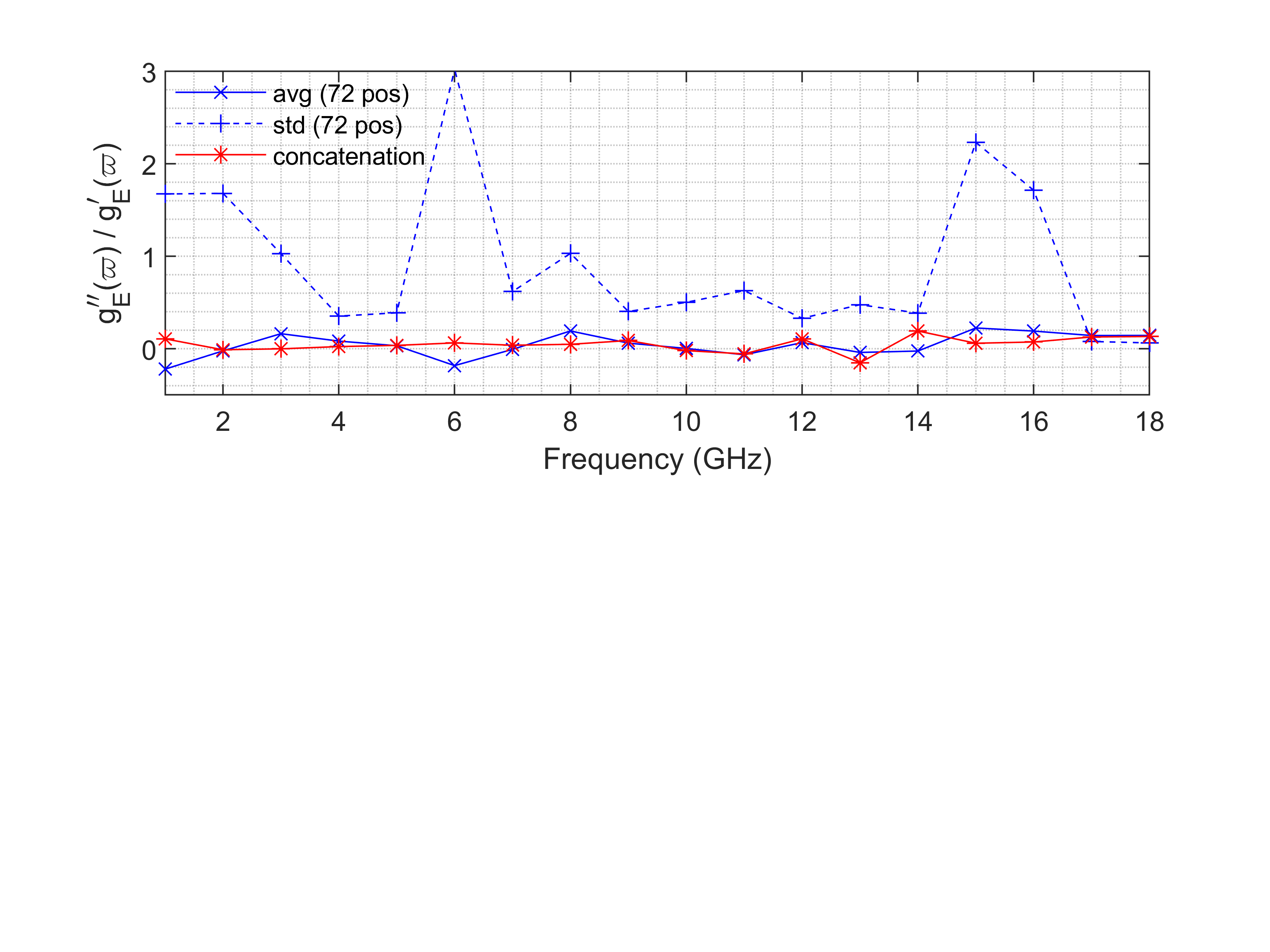}\\
\vspace{-4.3cm}\\
(b)\\
\hspace{-0.6cm}
\includegraphics[scale=0.65]{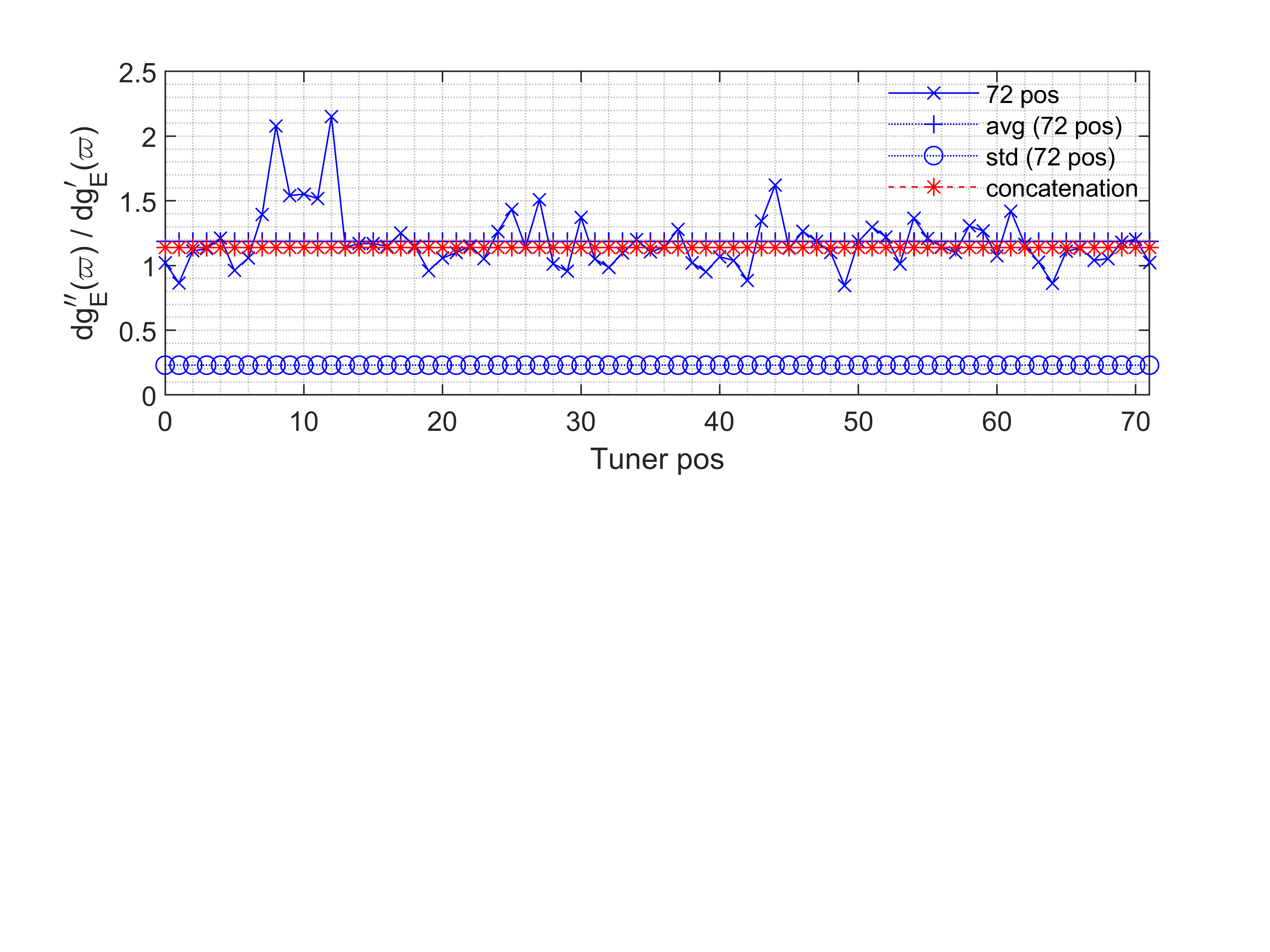}\\
\vspace{-4.3cm}\\
(c)\\
\hspace{-0.6cm}
\includegraphics[scale=0.65]{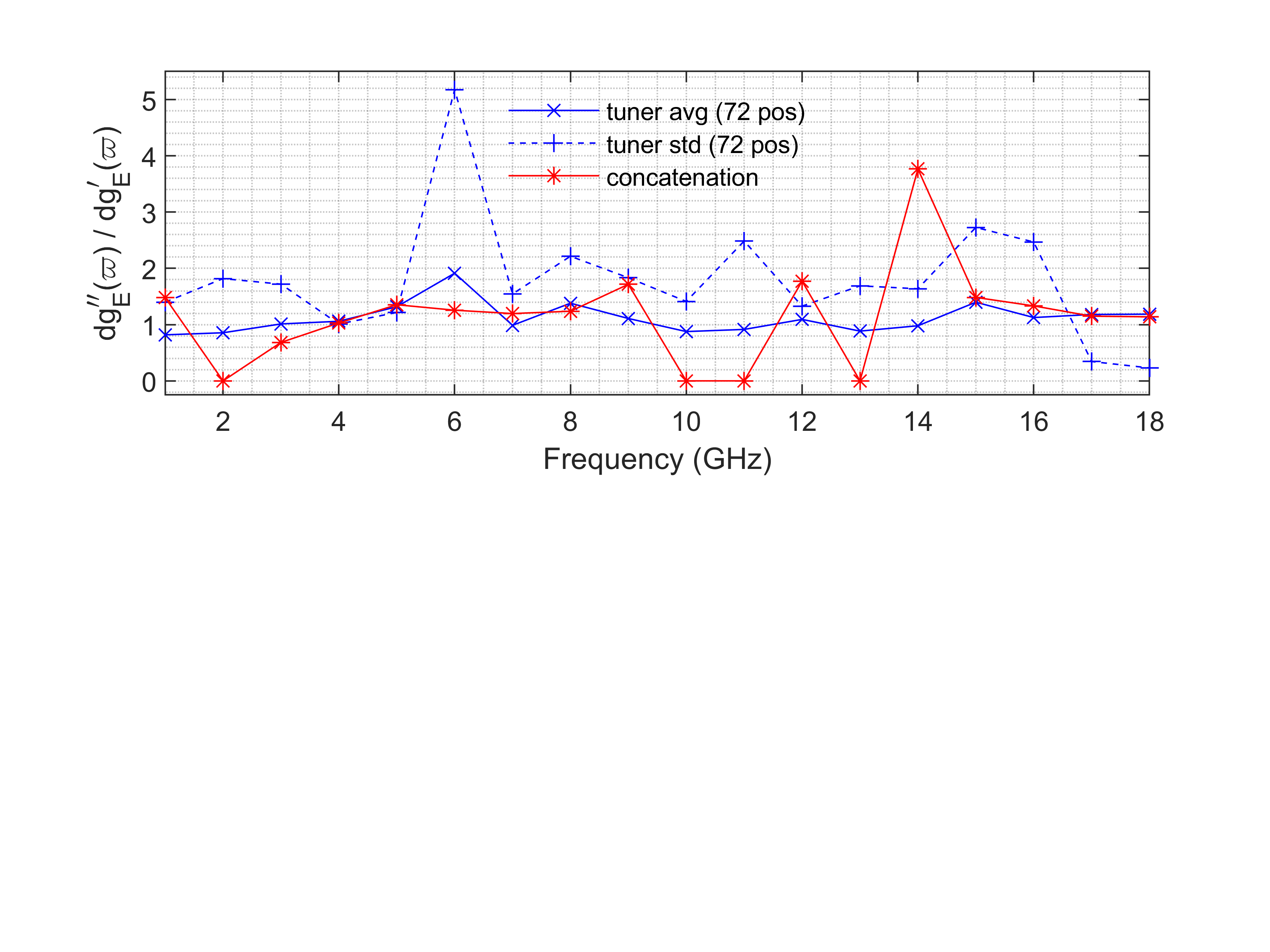}\\
\vspace{-4.3cm}\\
(d)\\
\end{tabular}
\end{center}
{
\caption{\label{fig:ratioslopesCSDFtoASDF}
\small
{
(a) CSDF-to-ASDF level ratio $g^{\prime\prime}_E(\varpi) / g^{\prime}_E(\varpi) = {6\sqrt{2\lambda^\prime_2}(\lambda^{\prime\prime}_1)^2 }/{\lambda^{\prime\prime}_3}$ at $f=18$ GHz for low $\varpi$ (mean: $0.144$; standard deviation: $0.063$);
(b) tune averaged mean and standard deviation of $g^{\prime\prime}_E(\varpi) / g^{\prime}_E(\varpi)$ as a function of $f$, compared with values for concatenated sweep.
(c) CSDF-to-ASDF slope ratio $\dot{g}^{\prime\prime}_E(\varpi)/\dot{g}^{\prime}_E(\varpi) = \sqrt{3\lambda^{\prime\prime}_1\lambda^\prime_2/\lambda^{\prime\prime}_3}$ at $f=18$ GHz (mean: $1.186$; standard deviation: $0.230$); 
(d) tune averaged mean and standard deviation of $\dot{g}^{\prime\prime}_E(\varpi)/\dot{g}^{\prime}_E(\varpi)$ as a function of $f$, compared with values for concatenated sweep.
}
}
}
\end{figure}

Fig. \ref{fig:ratioslopesCSDFtoASDF} compares relative levels and decay rates in the ASDF vs. CSDF at $f=18$ GHz.
The rate of the exponential decay of the CSDF, $\sqrt{\lambda^{\prime\prime}_3/(6\lambda^{\prime\prime}_1)} = 412.8$ rad/s, is close to the rate for the ASDF, $\sqrt{\lambda^\prime_2/2} = 421.1$ rad/s. This is no coincidence, as both rates arise as ratios of their first two Taylor expansion terms, for $\rho^{\prime\prime}_E(\tau)$ and $\rho^{\prime}_E(\tau)$, respectively. Their numerical similarity signifies that expanding both the ACF and CCF to the same order results in comparable levels of accuracy.
Consequently, the CSDF-to-ASDF slope ratios across tune states and CW frequencies in Fig. \ref{fig:ratioslopesCSDFtoASDF}(c) and (d) are typically of order $1$.

\section{Power Spectral Density}
As indicated in sec. \ref{sec:intro}, the experimental  evaluation of spectra may be performed for simpler scalar measurements, e.g., using an EMI receiver, spectrum analyzer or power meter. Since such instruments cannot resolve I- and Q-components, an assumption about their statistical relationship must be made, e.g., ideal I/Q circularity. Energy, power or field intensity then becomes the input quantity for extracting the SDFs, here referred to as the power SDF (PSDF), $g_U(\varpi)$. 
Since the input $U\propto |E|^2$ is real-valued, 
$g_U(\varpi)\equiv g^{\prime}_U(\varpi)$.

Fig. \ref{fig:psdf_complexE_and intensity} compares  the measured $g_U(\varpi)$ based on the EWK theorem vs. the periodogram method, while also comparing them to the ideal exponential ASDF $g^\prime_E(\varpi)$, formally equivalent to a zeroth-order PSDF \cite[eq. (77)]{arnaACFSDF_pt1}. The close agreement serves as a validation of the models in \cite[sec. V]{arnaACFSDF_pt1} and \cite{arnalocavg}. 

\section{ASDF and PSDF at Different CW Frequencies}

\subsection{Efficient Stirring}
Fig. \ref{fig:ASDFandPSDF_theovsmeas_SpinLarge_fvarious} compares the tune averaged ASDFs $\langle g^\prime_E (\varpi_k / \Delta \varpi) \rangle_{N_t}$ and PSDFs\footnote{In Fig. \ref{fig:ASDFandPSDF_theovsmeas_SpinLarge_fvarious}(b), the average PSDFs have not been normalized in order to separate and show the individual curves for different $f$ more clearly.} $\langle g_U (\varpi_k / \Delta \varpi) \rangle_{N_t}$ with the Pad\'{e} models developed in \cite{arnaACFSDF_pt1}, 
including LF and HF asymptotic approximations (dashed), 
based on the spectral moments that were evaluated in \cite{arnaACFSDF_pt2}.
Good agreement is 
found for the stir DC and Nyquist asymptotic dependencies and levels. 
From Fig. \ref{fig:ASDFandPSDF_theovsmeas_SpinLarge_fvarious}(a), it can be inferred that the relatively weak dependence of $\kappa^\prime$ on $f$ above 2 GHz \cite[Fig. 9(b)]{arnaACFSDF_pt2} follows from a compensation of the increasing LF peakedness and thinning of HF tails in the ASDF when $f$ is decreased.  
The predicted decrease of $\varpi^\prime_c$ with increasing $\kappa^\prime > 0$ \cite[eq. (44)]{arnaACFSDF_pt1} and with decreasing $f$ is confirmed by Fig. \ref{fig:ASDFandPSDF_theovsmeas_SpinLarge_fvarious}(a) and the decreasing $|\kappa^\prime(f)|$ in \cite[Fig. 9(b)]{arnaACFSDF_pt2}.
For lower $f$, an extension to higher-order ($m>2$) Pad\'{e} approximants involving additional deviation parameters $\kappa^{\prime(\prime)}_m$ \cite[eqs. (40) and (70)]{arnaACFSDF_pt1} allows for making further refinements of the estimated $\varpi^\prime_c$ and the asymptotic slopes and levels, particularly near $\varpi = 0$.
 
\begin{figure}[htb] 
\begin{center}
\begin{tabular}{c}
\hspace{-0.6cm}
\includegraphics[scale=0.65]{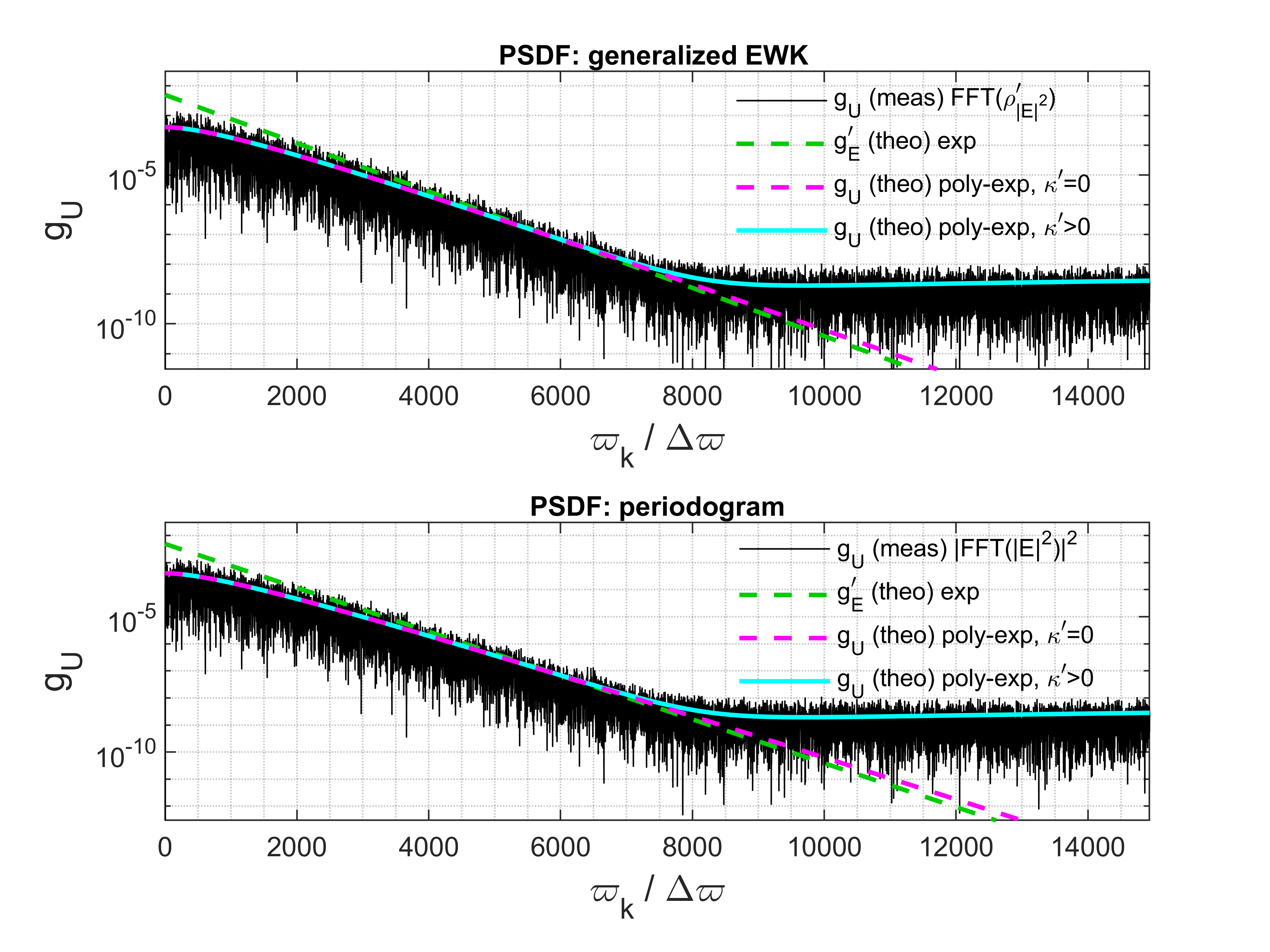}\\
\vspace{-0.9cm}\\
\end{tabular}
\end{center}
{
\caption{\label{fig:psdf_complexE_and intensity}
\small
{
Experimental PSDF $g_U(\varpi_k/\Delta\varpi)$ (black) from generalized EWK theorem-based (top) vs. periodogram-based (bottom) estimation, compared with theoretical exponential 0th-order 
(green), polynomial-exponential 1st-order 
(magenta) and 2nd-order (cyan) Pad\'{e} approximant models.
}
}
}
\end{figure}

\begin{figure}[htb] 
\begin{center}
\begin{tabular}{c}
\hspace{-0.6cm}
\includegraphics[scale=0.65]{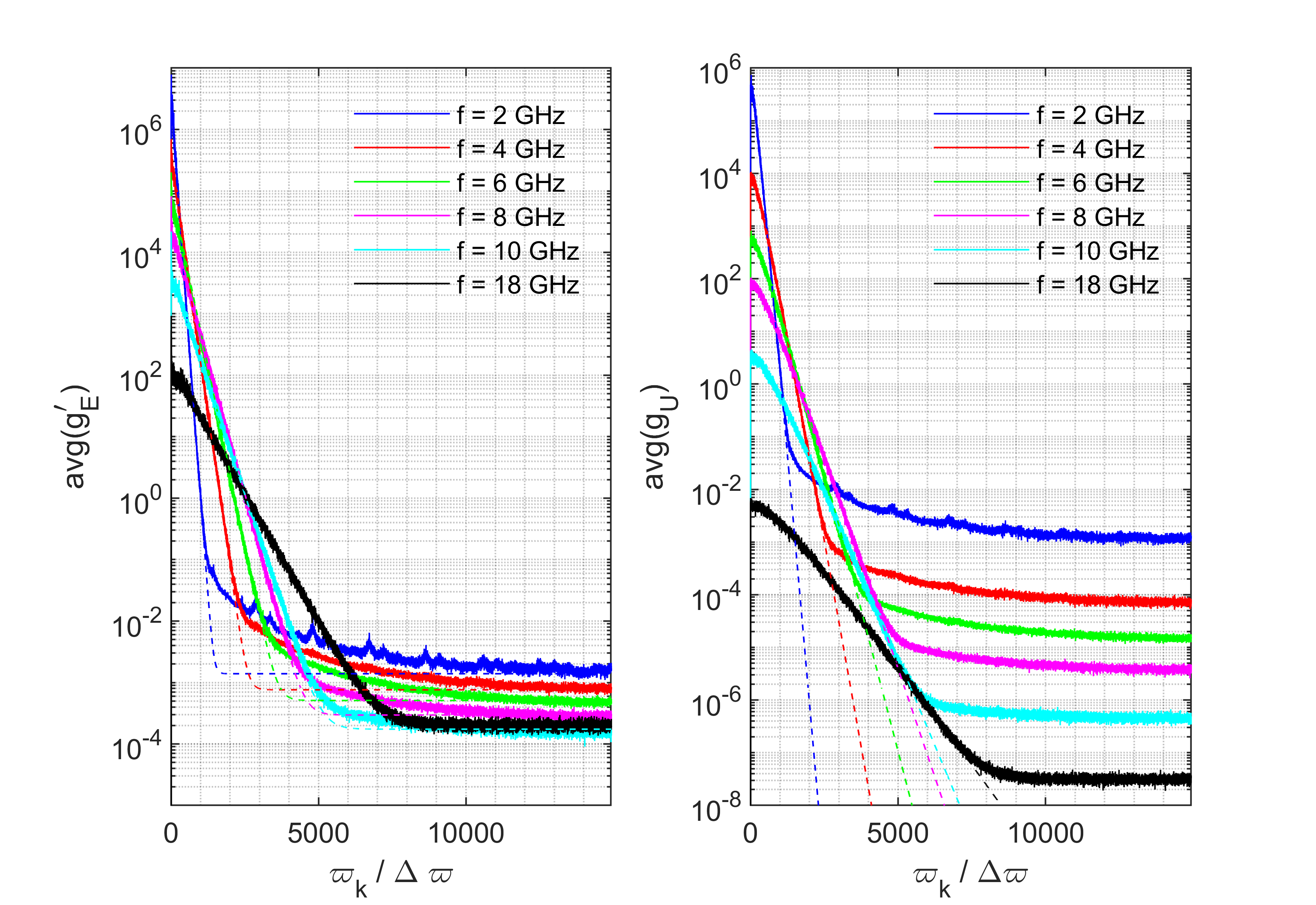}\\
\vspace{-0.9cm}\\
(a) ~~~~~~~~~~~~~~~~~~~~~~~~~~~~~~~(b)\\
\end{tabular}
\end{center}
{
\caption{\label{fig:ASDFandPSDF_theovsmeas_SpinLarge_fvarious}
\small
{
Comparison of (a) ASDFs and (b) PSDFs: Pad\'{e}-based models (dashed) vs. 72-tune average of measured SDFs (solid) at selected CW frequencies $f$, for stirring of large paddle and tuner averaging of small paddle. 
}
}
}
\end{figure}

The changing profile of $g^{\prime}_{E}(\varpi)$ with varying $f$ in Fig. \ref{fig:ASDFandPSDF_theovsmeas_SpinLarge_fvarious} originates from the similarity property of Fourier transforms,
\begin{align}
\cF\left [ {\rho}^{\prime(\prime)}_{E} \left (\tau/\tau^{\prime(\prime)}_c \right ) \right ](\varpi)
=
\tau^{\prime(\prime)}_c g^{\prime(\prime)}_{E} \left (\tau^{\prime(\prime)}_c \varpi \right )
\label{eq:Fourier_similaritytheorem}
\end{align}
in which the correlation time $\tau^{\prime(\prime)}_c$ serves as a scaling parameter.
In particular, if $f$ increases then $\tau^{\prime(\prime)}_c$ decreases, 
causing a contraction of $\rho^{\prime(\prime)}_{E}(\tau/\tau^{\prime(\prime)}_c)$ around $\tau=0$ for $\tau^{\prime(\prime)}_c > \Delta \tau$. In accordance with (\ref{eq:Fourier_similaritytheorem}), 
$g^{\prime(\prime)}_{E}(\varpi)$ then becomes stretched (dilated stir frequency) and flattened (lowered spectral amplitudes) as the SDF evolves towards white stir noise when $f \rightarrow +\infty$.

On the other hand, for the $n$th derivative of the CF, \begin{align}
\cF\left [ {\rho^{\prime(\prime)}_{E}}^{(n)}(\tau) \right ] (\varpi)
=
(-\rmj \varpi)^n g^{\prime(\prime)}_{E}(\varpi)
.
\label{eq:Fourier_deriiativetheorem}
\end{align}
In particular, $|\dot{\rho}^{\prime(\prime)}_{E}(|\tau| \rightarrow 0)| \sim \left. \varpi g^{\prime(\prime)}_{E}(\varpi)\right |_{\varpi \rightarrow \varpi_{\rm max}}$, signifying that the rate of CF decay at small lags relates to the HF values of the SDF. 
Hence, for mean-square differentiable $E(\tau)$, $g^{\prime(\prime)}_{E}(\varpi_{\rm max} \rightarrow +\infty) \rightarrow 0$ faster than $\varpi^{-1}$ when $\tau \rightarrow 0$.

\subsection{Inefficient Stirring: Interchange of Tuner and Stirrer Roles\label{sec:interchange}}
\subsubsection{Changes to SDFs}
Dual-stirred chambers have been investigated before to compare synchronized vs. interleaved tuning, using both paddles in conjunction \cite{mogl2010}. Here, interleaved stirring is used for comparing the stir performance of individual paddle wheels inside the {\em same\/} chamber. This eliminates any effect of the enclosure itself (geometry, volume). This constitutes a substitution method for assessing mode stirrers.

\begin{figure}[htb] 
\begin{center}
\begin{tabular}{c}
\hspace{-0.6cm}
\includegraphics[scale=0.65]{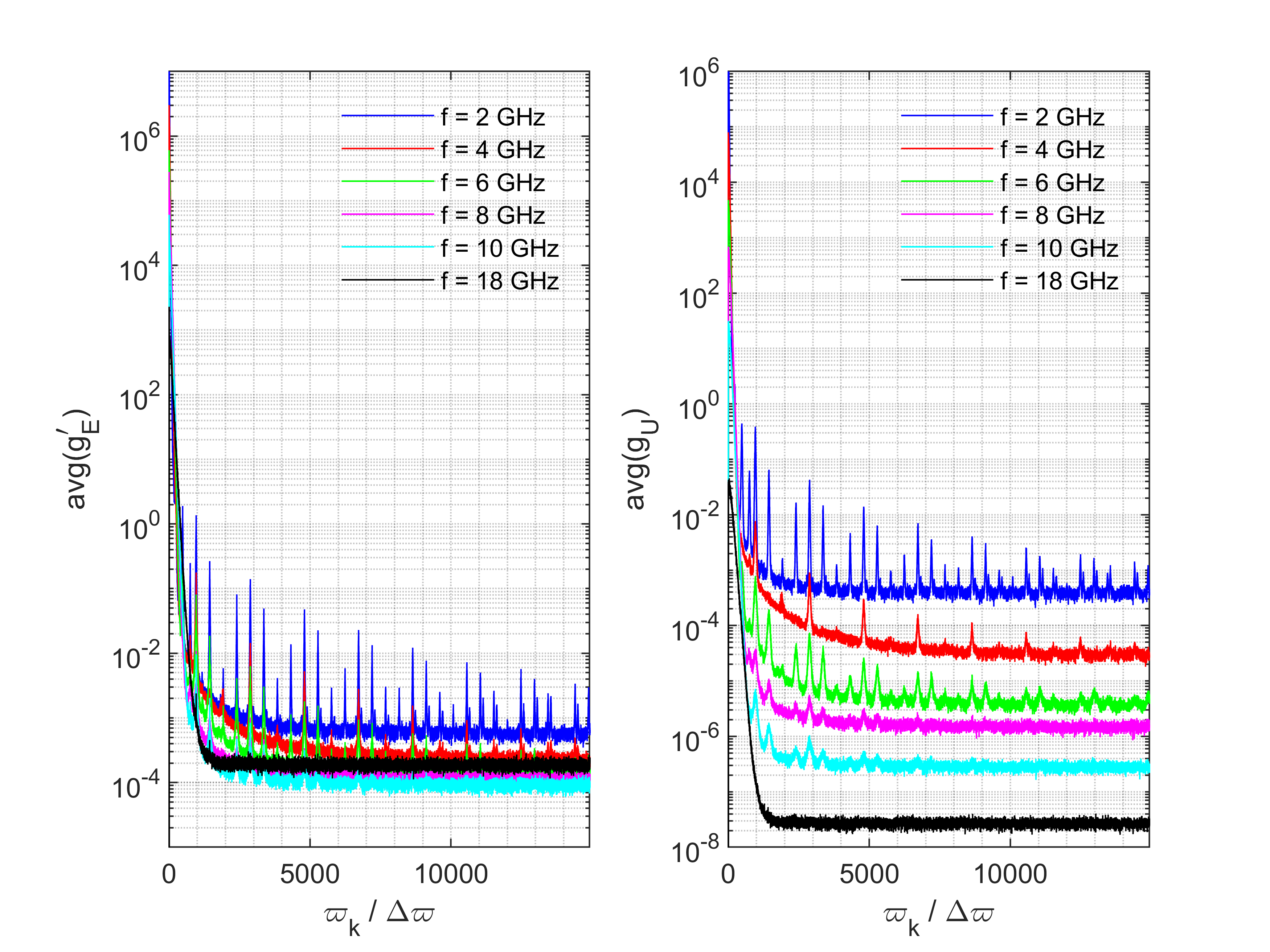}\\
\vspace{-0.9cm}\\
(a) ~~~~~~~~~~~~~~~~~~~~~~~~~~~~~~~(b)\\
\end{tabular}
\end{center}
\caption{\label{fig:ASDFandPSDF_meas_SpinSmall_fvarious}
\small{
Same as Fig. \ref{fig:ASDFandPSDF_theovsmeas_SpinLarge_fvarious} but for interchanged roles of small and large paddles, now as stirrer and secondary tuner, respectively. 
}
}
\end{figure}

Fig. \ref{fig:ASDFandPSDF_meas_SpinSmall_fvarious} shows SDFs obtained after interchanging the roles of stirrer and tuner, viz., now with mode stirring by the ``small'' tall narrow paddle and mode tuning by the ``large'' short wide paddle. 
A comparison with Fig. \ref{fig:ASDFandPSDF_theovsmeas_SpinLarge_fvarious} indicates a now reduced stir efficiency, manifesting itself through a steeper slope of the SDFs 
at low $\varpi$ because of a reduction of $\lambda^\prime_2$ (cf. \cite[eq. (46)]{arnaACFSDF_pt1}), {\it a fortiori} at high $f$,
and a lower $\varpi^\prime_c$ at all $f$. Correspondingly, in Fig. \ref{fig:ACF_meas_SpinSmall_f18GHz} an increase of $\tau^{\prime(\prime)}_c$ is witnessed, compared to Fig. \ref{fig:acf_ccf_Taylor_vs_Pade}. 
Note also that the stirrer's size and the excitation frequency are not fully interchangeable with respect to the assessment of stir efficiency, as can be witnessed from the sharper corner transition for $f=18$ GHz in Fig. \ref{fig:ASDFandPSDF_meas_SpinSmall_fvarious} compared to that for $f=2$ GHz in Fig. \ref{fig:ASDFandPSDF_theovsmeas_SpinLarge_fvarious}.

Earlier experimental \cite{lund2002} and theoretical \cite{arnadesign} investigations have indicated that wider paddles exhibit greater stirring efficiency than narrower ones. Unfortunately, the present results are unable to confirm this thesis, because of the large difference in the swept volume of both paddles.

\subsubsection{Stir Diagnostics}
Sharp periodic spikes are prominent in the SDFs for stirring by the small paddle in Figs. \ref{fig:ASDFandPSDF_meas_SpinSmall_fvarious} and \ref{fig:SweepAndASDF_meas_SpinSmall_f2GHz}(b), especially at low $f$, while being more subdued for stirring by the large paddle in Figs. \ref{fig:ASDFandPSDF_theovsmeas_SpinLarge_fvarious} and \ref{fig:SweepAndASDF_meas_SpinLarge_f2GHz}(b).
The spikes increase the value of $\lambda^\prime_4$ and hence $\kappa^\prime$. 
In Figs. \ref{fig:ASDFandPSDF_meas_SpinSmall_fvarious}(a) and \ref{fig:SweepAndASDF_meas_SpinSmall_f2GHz}(b), strong peaks occur at all integer multiples of the separation $\Delta k = 480\pm 1$. 
These represent quasi-harmonics of a fundamental at $f_{\rmEMI} = (480\pm 1)\Delta\varpi/(2\pi) \simeq 60.50\pm 0.13$ Hz, i.e., EMI traceable to the unfiltered mains power supply. 
Additional weaker LF sub- and interharmonics are observed at $k \equiv\varpi_k/\Delta\varpi \simeq 121$, $235$, $750$ and $1172$ but may have a small bias as they appear on the slope of the average ASDF.
Further HF interharmonics occur at $k_n \simeq 480 n + 100$, with a comparable magnitude as those of the mains harmonics.

In general, by evaluating the minimum and maximum detectable spacings in the stir frequency comb for a sampled harmonic EMI signal, viz., $2 \leq \Delta k \leq \lfloor N_s / 2 \rfloor$, the range of unambiguously detectable EMI harmonics follows as
\begin{align}
\Delta \varpi/\pi \leq f_{\rmEMI} \leq \lfloor N_s / 4 \rfloor \Delta\varpi / \pi
\label{ineq:fEMI}
\end{align} 
i.e., $2/T_s \leq f_{\rmEMI} \leq (2\Delta\tau)^{-1}$; here $0.252$ Hz $\leq f_{\rmEMI} \leq$ $1882$ Hz. The upper bound in (\ref{ineq:fEMI}) can be increased by several orders of magnitude when using a faster sampling VSA instead \cite{arnathresh}. 

\subsubsection{Stir Spectrogram}
The stir spectrogram in Fig. \ref{fig:SweepAndASDF_meas_SpinSmall_f2GHz}(c) reveals weak inhomogeneity through secondary tuning by the large paddle, e.g., at tune state $\ell \equiv \tau_{2,\ell}/\Delta \tau_2 = 17$ where $g^\prime_E(\varpi_k/\Delta\varpi)$ exhibits a lighter horizontal band of higher values, corresponding to a larger $\kappa^\prime$ \cite[Fig. 1]{arnaACFSDF_pt1} and hence less efficient stirring.
By contrast, more efficient stirring occurs, e.g., at $\ell  = 60$, where $g^\prime_E(\varpi_k/\Delta\varpi)$ exhibits a darker band. 

On the other hand, the pattern of uniform vertical bands indicates that the EMI spikes are non-intermittent here, but always occur at the same harmonics throughout the stirring and tuning processes, and that the tuning performance is uniform to a high degree. 
Also, the vertical bands differ in their width. 

More generally, the stir spectrogram can be used for detecting intermittent, i.e., temporary EMI through the appearance of interruptions (color changes) within vertical bands; e.g., because of uneven imperfections in the tuner's motor drive.  
For individual secondary tune states, spikes were found to exist at nearly all states except at $\ell=17$. Fig. \ref{fig:SweepAndASDF_meas_SpinSmall_f2GHz}(b) shows the ASDF for $\ell=1$. When $f$ is increased, any originally detectable EMI harmonic in the stir sweep becomes gradually overwhelmed by the increasing randomness of the field variation and hence by the increasing bandwidth of the stir process.

\begin{figure}[htb] 
\begin{center}
\begin{tabular}{c}
\hspace{-0.6cm}
\includegraphics[scale=0.65]{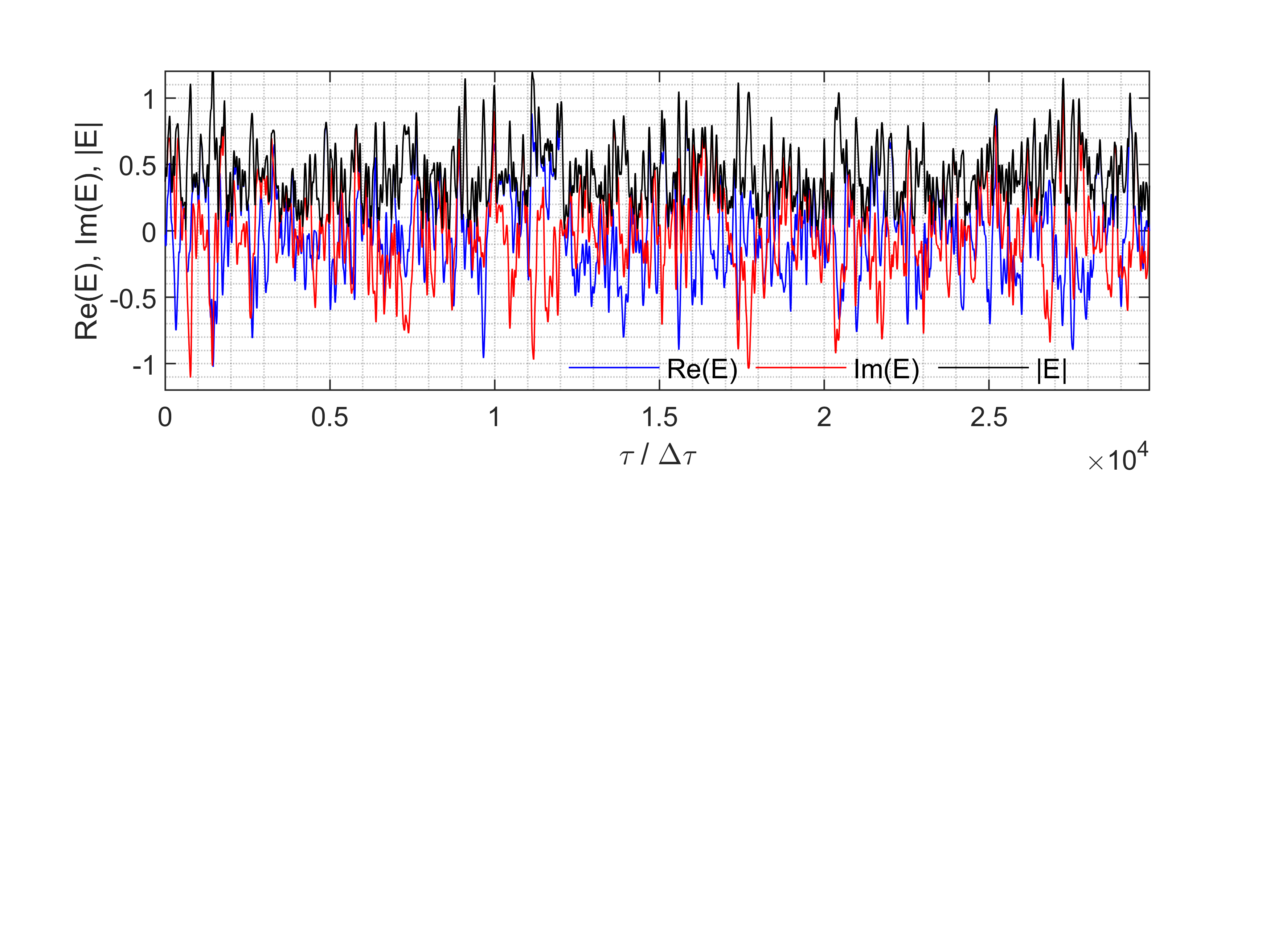}\\
\vspace{-4.5cm}\\
(a)\\
\hspace{-0.6cm}
\includegraphics[scale=0.65]{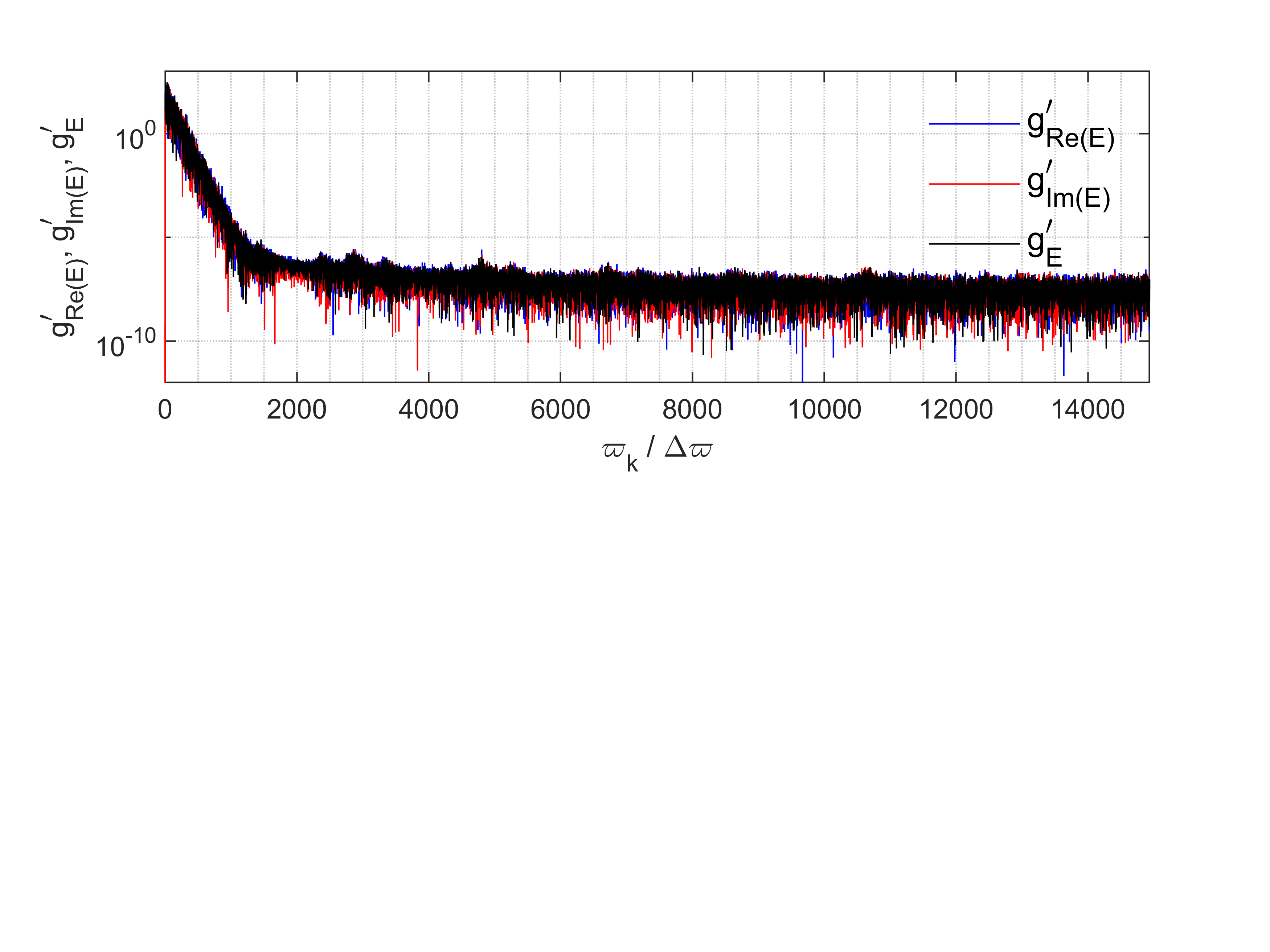}\\
\vspace{-4.5cm}\\
(b)
\end{tabular}
\end{center}
{
\caption{\label{fig:SweepAndASDF_meas_SpinLarge_f2GHz}
\small
{
(a) Stir sweeps and (b) associated ASDFs for large paddle stirring for arbitrary tune state of small paddle, at $f=2$ GHz.
}
}
}
\end{figure}

\begin{figure}[htb] 
\begin{center}
\begin{tabular}{c}
\hspace{-0.6cm}
\includegraphics[scale=0.65]{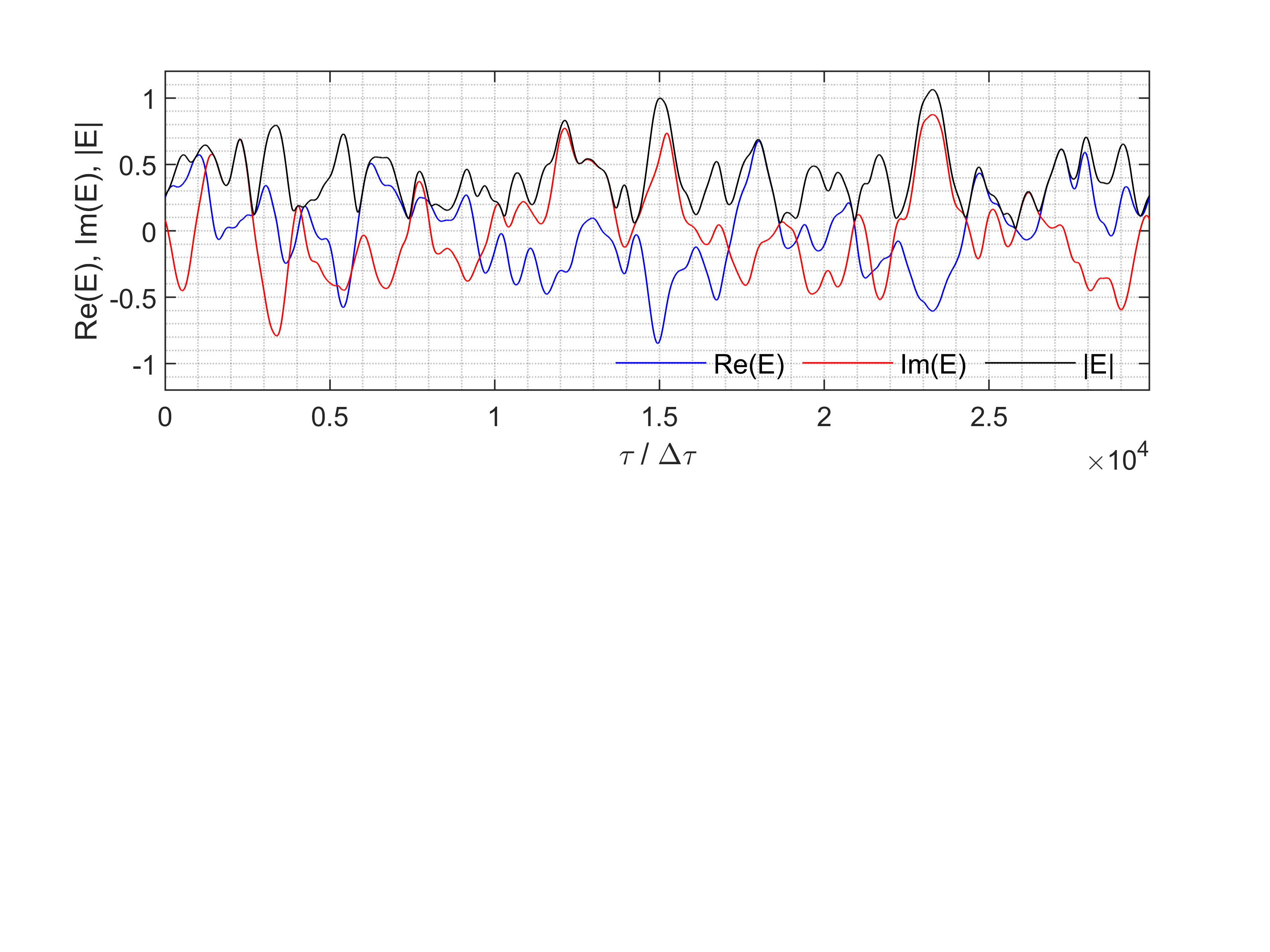}\\
\vspace{-4.4cm}\\
(a)\\
\hspace{-0.6cm}
\includegraphics[scale=0.65]{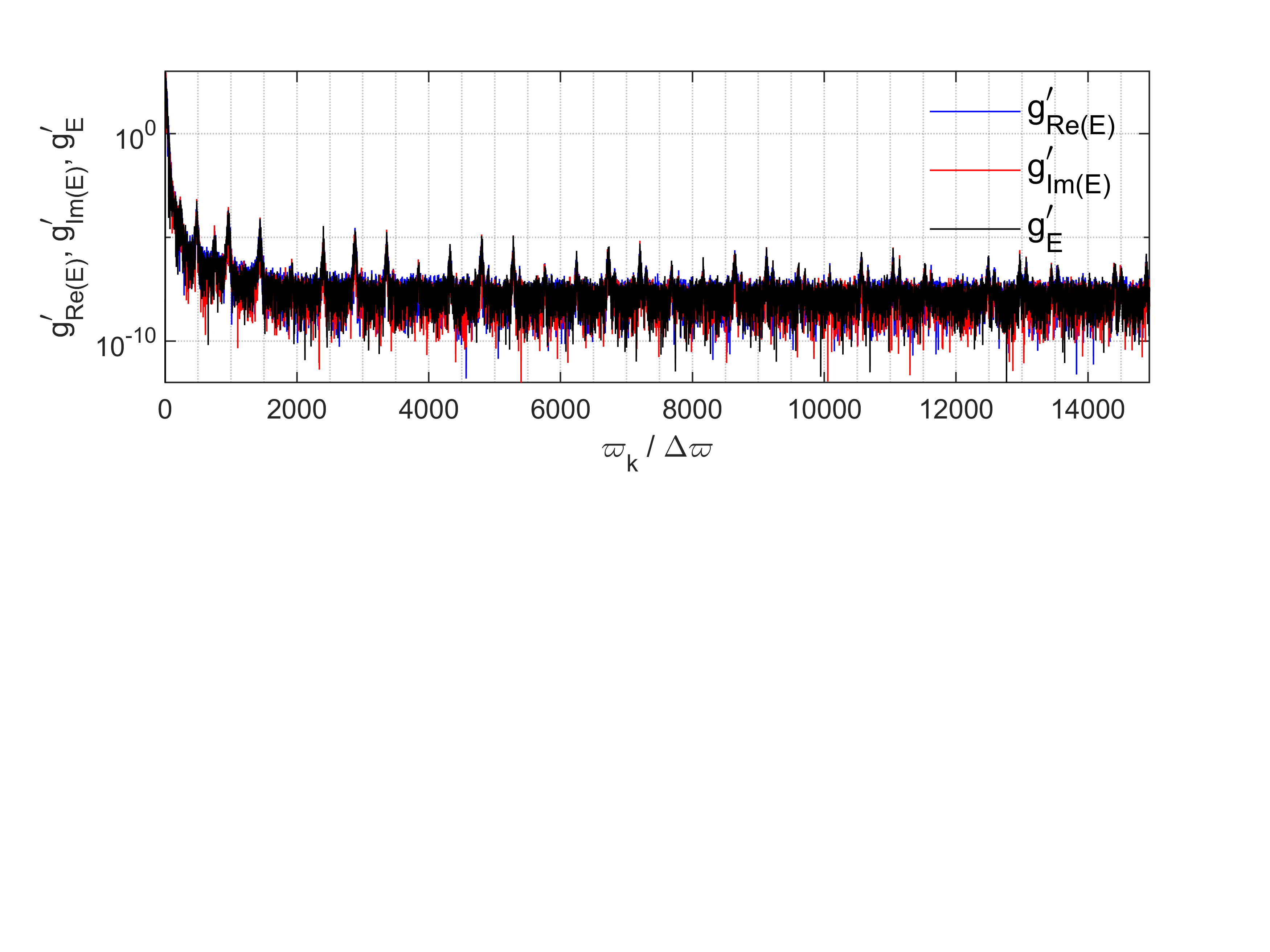}\\
\vspace{-4.4cm}\\
(b)\\
\hspace{-0.6cm}
\includegraphics[scale=0.65]{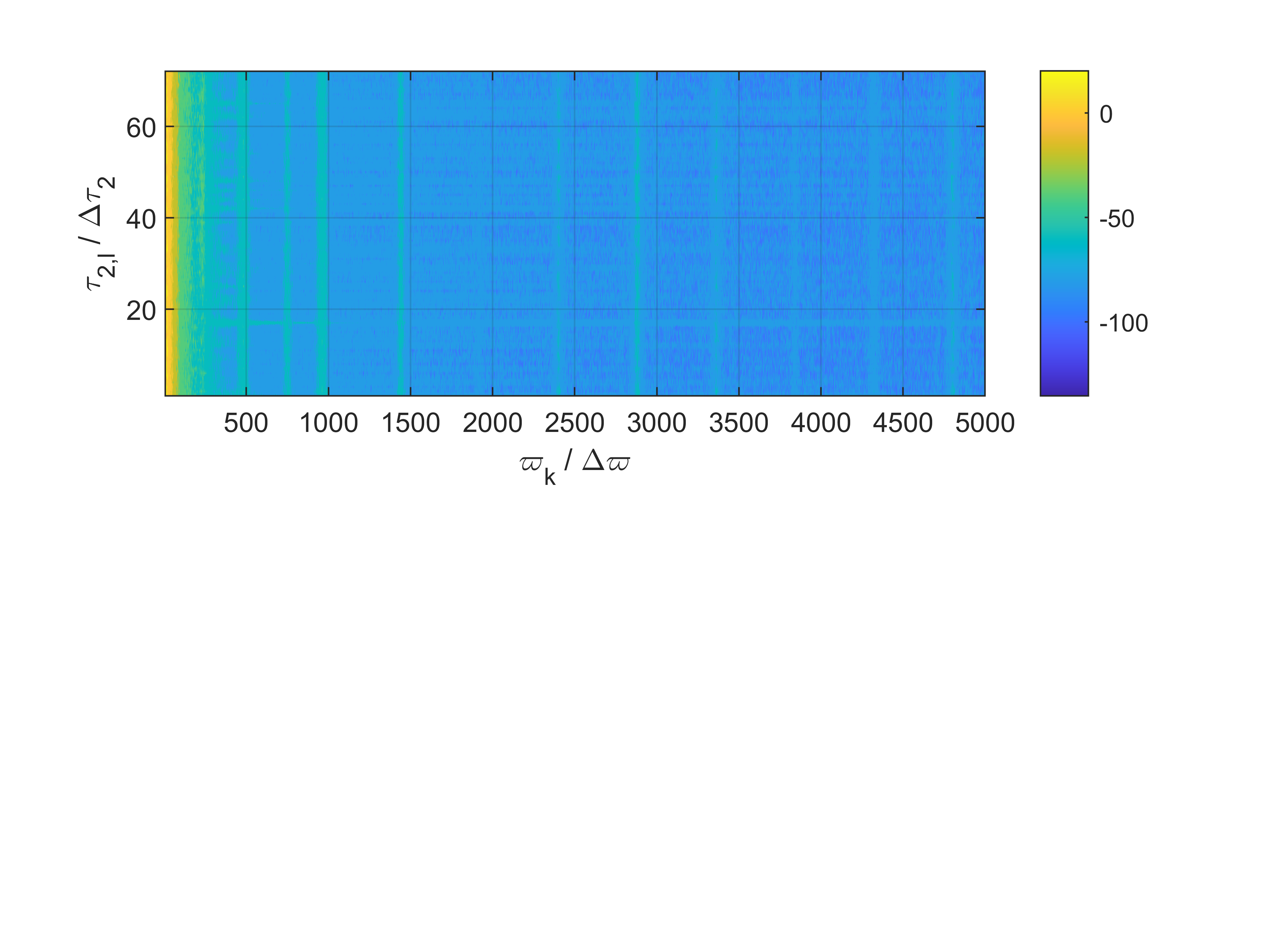}\\
\vspace{-4.4cm}\\
(c)\\
\end{tabular}
\end{center}
{
\caption{\label{fig:SweepAndASDF_meas_SpinSmall_f2GHz}
\small
{
(a)(b) Same as Fig. \ref{fig:SweepAndASDF_meas_SpinLarge_f2GHz} but for interchanged roles of now small paddle stirrer and large paddle tuner; 
(c) stir spectrogram of $g^\prime_E$ (in units dB) as a function of primary stir frequencies $5 \leq \varpi_k / \Delta \varpi \leq 5000$ ($\sim 0.00105 \leq \varpi_k \Delta\tau \leq 1.048$ rad) and secondary tune states $0 \leq \tau_{2,\ell} / \Delta\tau_2 \leq 71$ ($\sim 0$ $\leq \theta_\ell \leq 355$ deg).
}
}
}
\end{figure}

The EMI spikes in Figs. \ref{fig:ASDFandPSDF_theovsmeas_SpinLarge_fvarious}(a) and \ref{fig:SweepAndASDF_meas_SpinLarge_f2GHz}(b) are less pronounced than those in Fig. \ref{fig:ASDFandPSDF_meas_SpinSmall_fvarious}(a) and \ref{fig:SweepAndASDF_meas_SpinSmall_f2GHz}(b) because a larger stirrer produces higher and flatter SDF levels. Nevertheless, they still appear with the same $\Delta k\simeq 480$.
Thus, inefficient stirring detects this EMI more easily.
Observe that, at $f=2$ GHz, the locations of (weak) LF local maxima across stir states in Figs. \ref{fig:ASDFandPSDF_theovsmeas_SpinLarge_fvarious}(a) and \ref{fig:SweepAndASDF_meas_SpinLarge_f2GHz}(b) correspond to those for local minima in $\nu_{g^\prime_E}(\varpi_k\Delta\tau)$ in Fig. \ref{fig:CffVarSDF_SpinLarge_f18GHz} across tune states. 
This confirms the common source of EMI affecting both the stirrer and the tuner. 

\begin{figure}[htb] 
\begin{center}
\begin{tabular}{c}
\vspace{-0.5cm}\\
\hspace{-0.6cm}
\includegraphics[scale=0.65]{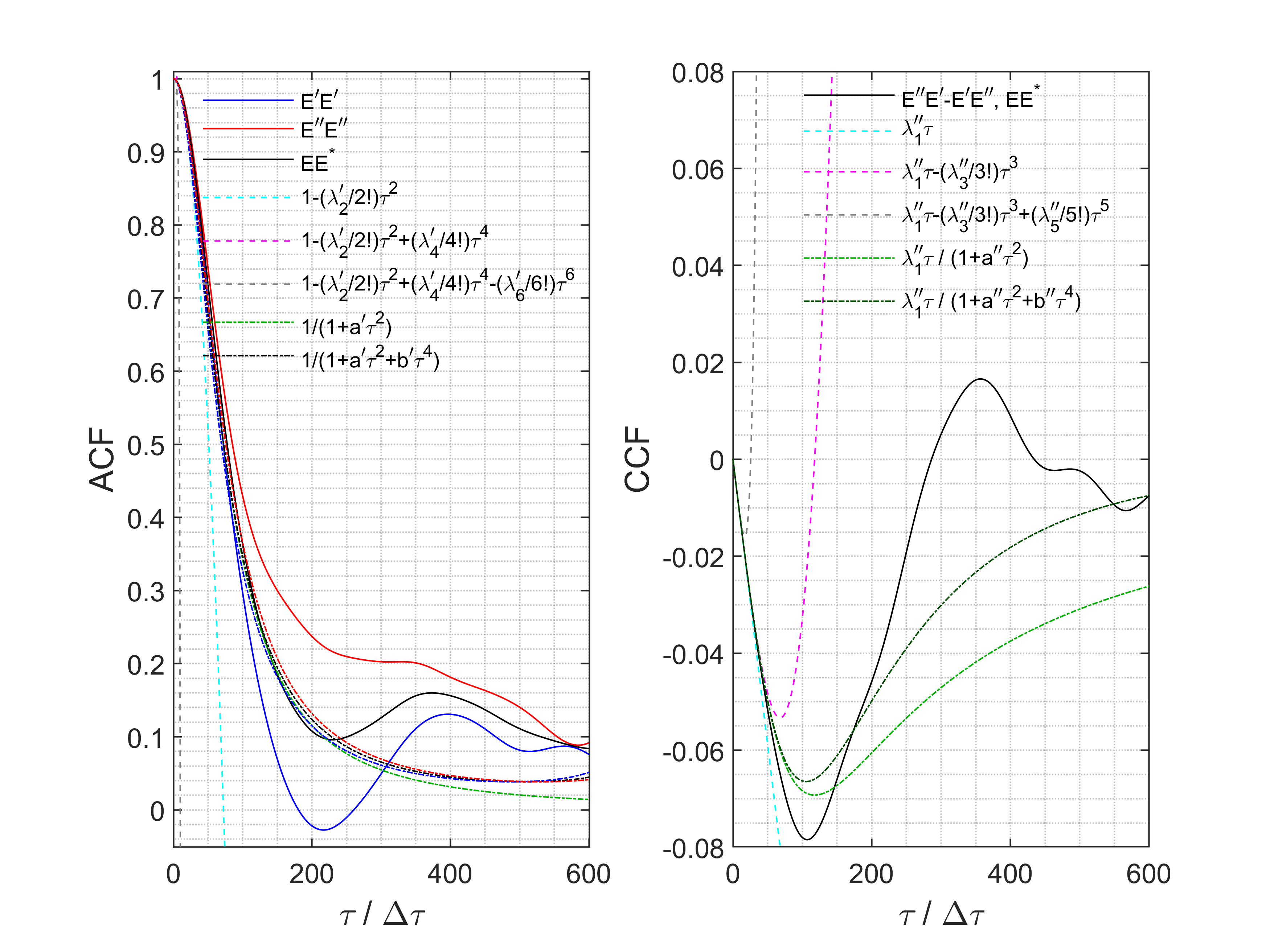}\\
\vspace{-0.9cm}\\
(a)~ ~~~~~~~~~~~~~~~~~~~~~~~~~~~(b)\\
\end{tabular}
\end{center}
{
\caption{\label{fig:ACF_meas_SpinSmall_f18GHz}
\small
{
(a) ACFs and (b) CCFs as in Fig. \ref{fig:acf_ccf_Taylor_vs_Pade} but for interchanged roles of small and large paddle as stirrer and tuner, at $f=18$ GHz. 
}
}
}
\end{figure}

Notch or wavelet filtering of harmonics 
improves the accuracy of the [0/2]-order Pad\'{e} approximant and its ASDF, although the first-order approximant prior to any filtering (shown in green in Fig. \ref{fig:ACF_meas_SpinSmall_f18GHz}) offers an adequate model already.

\section{Conclusion}
In this part III, experimental CFs and SDFs were compared with theoretical Pad\'{e}-based models \cite{arnaACFSDF_pt1}, showing good agreement. Periodogram- and generalized EWK-based methods produce near-identical results (Fig.  \ref{fig:psdf_complexE_and intensity}).
First- and second-order Pad\'{e} approximants for stir ACFs and CCFs were found to outperform Taylor approximations of the same and higher order. 
A simple first-order (one-parameter, $\lambda^\prime_2$) Pad\'{e} model already offers highly accurate ACF values down to $\rho^\prime_E(\tau) \sim 0.1\ldots 0.3$, with a further extended range as the efficiency of the stirrer increases (Figs. \ref{fig:acf_ccf_Taylor_vs_Pade} and \ref{fig:ACF_meas_SpinSmall_f18GHz}).

Extensions to higher-order Pad\'{e} approximants, offering improved accuracy and/or an extended domain or range for stir lags or frequencies, are possible at the expense of solving cubic or higher-order biquadratic root equations for the parameters of the CF model, extending \cite[eqs. (25) and (61)]{arnaACFSDF_pt1}. 
However, such approximants depend on higher-order spectral moments $\lambda^{\prime(\prime)}_m$ in their coefficients. The accurate estimation of $\lambda^{\prime(\prime)}_m$ for $m \gg 1$ from discrete finite-difference stir sweep data or their ACF is prone to rapidly increasing bias and uncertainty with increasing $m$ \cite{arnaACFSDF_pt2}.
This issue becomes already apparent in the second-order Pad\'{e}-based model for the ACF, where certain asymptotic spectral features exhibit heightened sensitivity to the spectral kurtosis $\kappa^\prime$ (Tbl. I). 

Since the DC-to-Nyquist drop of the ASDF level for $\kappa^\prime > 0$ is $1/\sqrt{\kappa^{\prime}}$ \cite[eq. (43)]{arnaACFSDF_pt1}, i.e., not explicitly dependent on $\lambda^\prime_2$, this interpretation offers a practical method of estimating $\kappa^{\prime}$ directly from the experimental ASDF itself (Fig. \ref{fig:fsdf_WK_vs_periodogram}),
thus bypassing the difficulties in the estimation of  individual $\lambda^\prime_m$ \cite[eq. (27)]{arnaACFSDF_pt1}.

The shallower the negative slope $-(2/\lambda^\prime_2)(1-2\kappa^\prime)$ for the ASDF's envelope in going from stir DC toward $\varpi^\prime_c$ \cite[eq. (47)]{arnaACFSDF_pt1}, the greater the stir efficiency. Ideal stirring  corresponds to pure white stir noise ($\lambda^\prime_2\rightarrow +\infty$).

The combined use of one stirrer with one tuner offers a capability of generating an ensemble of one-dimensional ACFs and SDFs. 
This allows for estimating the average and the RMS spread of the autocovariance or autocorrelation at an arbitrary stir lag $\tau$, and the spectral density at a particular stir frequency $\varpi$ (Fig. \ref{fig:CffVarSDF_SpinLarge_f18GHz}). 
When the number of tune states $N_t$ is increased, the ensemble averaged SDF of the stirrer exhibits a reduced spread of fluctuations across its entire stir band; notably also compared to that for a concatenation of stir sweeps. For the latter, increasing the number of data points to $N_t N_s$ is unable to reduce the spectral spread \cite{jenk1968} (Fig. \ref{fig:ASDF_SpinLarge_concat_f18GHz}). 
Ensemble averaging of SDFs avoids the need for low-pass filtering or Welch's method on a single-trace SDF. This eliminates nonuniform spectral biasing across the stir spectrum, which distorts not only the estimated SDF \cite{prie1981}, \cite{jenk1968} but also affects the SDF-based estimated spectral moments. This distortion is particularly strong near $\varpi_{\rm max}$.

Finally, at sufficiently low excitation frequencies, the stir SDF was shown to permit performing diagnostics, e.g., identification of intermittent or permanent EMI (Figs. \ref{fig:ASDFandPSDF_theovsmeas_SpinLarge_fvarious}--\ref{fig:SweepAndASDF_meas_SpinSmall_f2GHz}), in particular using the stir spectrogram (Fig. \ref{fig:SweepAndASDF_meas_SpinSmall_f2GHz}(c)).



\begin{thebibliography}{9}
\bibitem{arnaACFSDF_pt1} L. R. Arnaut, ``Correlation and spectral density functions in mode-stirred reverberation -- I. Theory,'' \it IEEE Trans. Electromagn. Compat., \rm 2024, doi 10.1109/TEMC.2024.3355801.
\bibitem{arnaACFSDF_pt2} L. R. Arnaut and J. M. Ladbury, ``Correlation and spectral density functions in mode-stirred reverberation -- II. Spectral moments, sampling, noise, EMI and understirring,'' \it IEEE Trans. Electromagn. Compat., \rm 2024, doi 10.1109/TEMC.2024.3375280.
\bibitem{arnalocavg} L. R. Arnaut, ``Effect of local stir and spatial averaging on the measurement and testing in mode-tuned and mode-stirred reverberation chambers,'' \it IEEE Trans. Electromagn. Compat., \rm vol. 43, no. 3, pp. 305--325, Aug. 2001.
\bibitem{arnathresh} L. R. Arnaut, ``Threshold level crossings, excursions, and extrema in immunity and fading testing using multistirred reverberation chambers,'' \it IEEE Trans. Electromagn. Compat., \rm vol. 62, no. 5, pp. 1638--1650, Oct. 2020.
\bibitem{prie1981} M. B. Priestley, \it Spectral Analysis and Time Series -- Vol. 1: Univariate Series. \rm Academic Press: London, U.K., 1981.
\bibitem{jenk1968} G. M. Jenkins and D. G. Watts, \it Spectral Analysis and Its Applications. \rm Holden-Day: San Francisco, CA, 1968.
\bibitem{mogl2010} F. Moglie and V. Mariani Primiani, ``Evaluation of uncorrelation and statistics inside a reverberation chamber in presence of two independent stirrers,'' \it Proc. 2010 IEEE Symp Electromagn. Compat., \rm Ft. Lauderdale, FL, pp. 515--519, Jul. 2010. 
\bibitem{lund2002} O. Lund\'{e}n and M. B\"{a}ckstr\"{o}m, ``Design of experiment. How to improve reverberation chamber mode-stirrer efficiency,'' \it FOI Tech. Rep. FOI-R-0468-SE, \rm FOI, Link\"{o}ping, Sweden, Apr. 2002.
\bibitem{arnadesign} L. R. Arnaut, ``Effect of size, orientation, and eccentricity of mode stirrers on their performance in reverberation chambers,'' \it IEEE Trans. Electromagn. Compat., \rm vol. 48, no. 3, pp. 600--602, Aug. 2006.
\end{thebibliography}
\end{document}